\documentclass{article}

\usepackage{arxiv}

\usepackage[utf8]{inputenc} 
\usepackage[T1]{fontenc}    
\usepackage{hyperref}       
\usepackage{url}            
\usepackage{booktabs}       
\usepackage{amsfonts}       
\usepackage{nicefrac}       
\usepackage{microtype}      
\usepackage{lipsum}
\usepackage{graphicx}
\graphicspath{ {./images/} }
\usepackage{natbib}
\usepackage{multirow}
\usepackage{caption}
\usepackage{array}
\usepackage{tabularx} 
\usepackage{ragged2e} 
\usepackage{threeparttable} 
\usepackage{amsmath}

\title{Improving Medium Range Severe Weather Prediction through Transformer Post-processing of AI Weather Forecasts}

\author{
  Zhanxiang Hua* \\
  Department of Atmospheric and Climate Science \\
  University of Washington \\
  Seattle, WA \\
  \texttt{zxhua@uw.edu} \\
  \And
  Ryan A. Sobash \\
  US NSF National Center for Atmospheric Research \\
  Boulder, CO \\
  \texttt{sobash@ucar.edu} \\ 
  \And
  David John Gagne II \\
  US NSF National Center for Atmospheric Research \\
  Boulder, CO \\
  \texttt{dgagne@ucar.edu} \\ 
  \And
  Yingkai Sha \\
  US NSF National Center for Atmospheric Research \\
  Boulder, CO \\
  \texttt{ksha@ucar.edu} \\ 
  \And
  Alexandra Anderson-Frey \\
  Department of Atmospheric and Climate Science \\
  University of Washington \\
  Seattle, WA \\
  \texttt{akaf@uw.edu} \\ 
}

\begin{document}
\maketitle
\begin{abstract}
Improving the skill of medium-range (3-8 day) severe weather prediction is crucial for mitigating societal impacts. This study introduces a novel approach leveraging decoder-only transformer networks to post-process AI-based weather forecasts, specifically from the Pangu-Weather model, for improved severe weather guidance. Unlike traditional post-processing methods that use a dense neural network to predict the probability of severe weather using discrete forecast samples, our method treats forecast lead times as sequential ``tokens'', enabling the transformer to learn complex temporal relationships within the evolving atmospheric state. We compare this approach against post-processing of the Global Forecast System (GFS) using both a traditional dense neural network and our transformer, as well as configurations that exclude convective parameters to fairly evaluate the impact of using the Pangu-Weather AI model. Results demonstrate that the transformer-based post-processing significantly enhances forecast skill compared to dense neural networks. Furthermore, AI-driven forecasts, particularly Pangu-Weather initialized from high resolution analysis, exhibit superior performance to GFS in the medium-range, even without explicit convective parameters. Our approach offers improved accuracy, and reliability, which also provides interpretability through feature attribution analysis, advancing medium-range severe weather prediction capabilities. 
\end{abstract}


\section{Introduction}
Severe convective weather hazards (i.e., tornadoes, hail $\geq$ 1" in diameter, and convectively-generated wind gusts $\geq$ 50 knots), pose significant threats to life and property across the contiguous United States (CONUS). Accurate and timely prediction of these events is crucial for mitigating their impacts. The characteristics of the pre-convective environment play a critical role in determining the likelihood and intensity of severe weather. Many studies have focused on identifying the key distinguishing features of severe convective environments \citep{Thompson2003-co,Thompson2013-wz,shield2022diagnosing, hua2022self, hua2023tornadic}. These studies highlight the complexity and nuanced nature of the atmospheric conditions that lead to severe weather.

Traditionally, severe weather forecasting has relied heavily on Numerical Weather Prediction (NWP) models, which provide both environmental and storm-scale information that can be used to infer the likelihood of convective hazards in the future. This includes the use of convection-allowing models (CAMs) at lead-times of 1--2 days (e.g., the High-Resolution Rapid Refresh model \citep{Dowell2022}) and coarser NWP ensembles at lead times of up to and beyond a week (e.g., the Global Ensemble Forecasting System). Since these NWP systems do not explicitly represent convective hazards, machine learning (ML) can be used to post-process NWP output \citep{vannitsem2021statistical}, using predicted environmental and storm-scale fields to generate hazard probabilities \citep{mcgovern2023}. ML models such as random forests \citep{hill2020,hill2023,loken2020,loken2022}, dense neural networks \citep{Sobash2020-zs,SobashAhijevych2024,SobashAhijevych2025}, and convolutional neural networks \citep{Gagne2019-it,Shaetal2024} have all been used to generate skillful guidance for convective hazards over CONUS using operational NWP model output at lead-times from hours to days. Beyond convective hazards, several works have focused on postprocessing forecasts for extreme events such as high-wind scenarios \citep{veldkamp2021statistical, schulz2022machine}.

This study extends this line of research by introducing and evaluating two novel approaches to severe weather prediction: 1) a post-processing system to generate convective hazard probabilities based on the use of transformer networks, rather than random forests or dense neural networks as in prior work, and 2) the use of AI NWP emulators to provide forecasts of environmental conditions. AI NWP emulators, like Pangu-Weather \citep{bi2023accurate}, offer a significant advantage in terms of computational speed and have demonstrated comparable or even superior performance to operational global NWP forecasts \citep{rasp2023weatherbench}. Recent work has also begun to explore the post-processing of AI-based weather model output for various applications, such as ensemble postprocessing for uncertainty quantification \citep{buelte24} and forecast refinement \citep{npg-31-247-2024}. These studies also test the ability of such models to reliably simulate other hazards like tropical cyclones \citep{demaria2024evaluation}, heatwaves \citep{meng2025deep}, and to reconstruct derecho climatology \citep{li2024derecho} from postprocessed reanalysis and observations. A recent study by \cite{Hua2025performance} shows that Pangu-Weather is more accurate than GFS in forecasting wind shear and storm-relative helicity at the grid point and hour closest to the tornado report at a one-day lead time. We hypothesize that gains in large-scale forecast skill will translate into more skillful convective hazard forecasts, especially in the medium-range (i.e., lead-times of 3--8 days) where the skill differences between traditional and data-driven NWP are most significant.

To generate hazard probabilities, we post-process the AI NWP forecast output using decoder-only transformers. To our knowledge, this is the first study to use transformers for severe weather prediction, although other recent papers have proposed transformer-based post-processing models \citep{bouallegue2024improving, van2025self} applied to forecast fields. Taking inspiration from large language modeling, we treat each forecast lead time as a "token" within a sequence. Rather than predicting the next token in the sequence, our model predicts the probability of at least one severe weather report (any hazard) within a 40 km radius for 24-hour periods from day 1 through day 8. The transformer architecture allows the model to learn complex temporal relationships and dependencies within the sequence of input forecasts, effectively capturing the evolution of atmospheric conditions leading to severe weather.

In this work, we document a set of experiments to test the impact of the use of AI NWP emulators and transformers for convective hazard prediction by comparing convective hazard forecast skill to existing baselines during 2024, an active severe weather season across CONUS. Weather forecasts are generated using Pangu-Weather initialized with several different initial conditions and are compared to forecasts from the Global Forecast System (GFS). In addition, the forecasts are post-processed to generate hazard forecasts, and the transformer-based post-processing system is compared against dense neural networks. We also explore combining the data-driven AI and traditional NWP forecasts to better capture forecast uncertainty in an ensemble approach. Together, these experiments provide insight into the ability of AI NWP emulators to improve forecasts of small-scale hazards and extend practical predictability of convective hazards into the medium-range.

\section{Methodology}\label{sec: methodology}

\subsection{Preprocessing Global NWP and AI Forecast Dataset}
In this work, post-processed convective hazard forecasts are generated from two different deterministic NWP systems: the operational GFS and Pangu-Weather, an AI NWP model. The GFS \citep{yang2006evaluation}, which represents the baseline for traditional NWP forecasting in this study, provides comprehensive atmospheric forecasts globally on a horizontal grid having 13 km grid spacing and 127 vertical levels. The initial conditions of the GFS forecast are produced by the Global Data Assimilation System, which runs a 4D hybrid ensemble-variational data assimilation scheme. The system initiates forecasts four times daily at 00, 06, 12, and 18 UTC, extending forecasts up to 16 days using the Finite-Volume Cubed-Sphere (FV3) model coupled with the NCEP Global Wave Model; here we only use the 00 UTC initialized forecasts. Forecast grids are archived on a 0.25 by 0.25-degree global latitude-longitude grid, offering 3-hourly forecast intervals up to 240 hours. 

The Pangu-Weather AI NWP model emulator \citep{bi2023accurate} leverages a 3D Earth-specific transformer with hierarchical temporal aggregation for fast and accurate global weather forecasting. Trained on 1979–2017 ERA5 \citep{hersbach2020era5} reanalysis data with a resolution of 0.25 by 0.25-degree, four separate networks are used to predict at four different lead times (1, 3, 6, and 24 hours) for mean sea level pressure, surface temperature, wind, and 5 pressure-level variables. The 5 pressure-level variables are geopotential height, temperature, specific humidity, and U and V components of wind at 13 pressure levels. The set of variables used as input to the model is identical to the set of variables it predicts as output, enabling iterative prediction. Pangu-Weather employs a hierarchical temporal aggregation approach, which is a greedy algorithm designed to forecast in the fewest possible steps. For example, to generate a 54-hour lead time forecast, Pangu-Weather iteratively applies the 24-hour model twice, followed by the 6-hour model. We generate and save 6-hourly output for all Pangu-Weather forecasts.

We use Pangu-Weather to generate daily 00 UTC-initialized forecasts from 1 January 2018 -- 1 August 2024 with three different initial conditions (ICs): the ERA5 reanalysis data, the ECMWF High-resolution (HRES) analysis (used to initialize the ECMWF Integrated Forecasting System), and GFS analysis. The HRES analysis has a nominal grid point spacing of 9 km and was bilinearly interpolated to match the grid point spacing of ERA5 reanalysis using xESMF \citep{zhuang2019xesmf}. It is important to note that Pangu-Weather was trained with ERA5 and has not been fine-tuned on either HRES or GFS analyses. Using HRES and GFS for ICs allows us to assess its potential performance using operational datasets, since ERA5 reanalysis data is not available in real-time.

\begin{table}
\centering
\small
\begin{tabularx}{\textwidth}{ >{\RaggedRight}X m{3em} m{3em} m{3em} m{3em} m{8em} }
 \hline\hline
 Variable & GFS & Pangu-ERA5 & Pangu-GFS & Pangu-HRES & GFS+Convect  \\
\hline
Latitude, Longitude, Forecast Day & \checkmark & \checkmark & \checkmark & \checkmark & \checkmark \\
Day of Year (encoded with sine and cosine) & \checkmark & \checkmark & \checkmark & \checkmark & \checkmark  \\
Local Solar Hour (encoded with sine and cosine) & \checkmark & \checkmark & \checkmark & \checkmark & \checkmark  \\

2-m Temperature & \checkmark & \checkmark & \checkmark & \checkmark & \checkmark \\
Mean sea level pressure & \checkmark & \checkmark & \checkmark & \checkmark & \checkmark \\
10-m wind speed & \checkmark & \checkmark & \checkmark & \checkmark & \checkmark \\
700--500 hPa lapse rate & \checkmark & \checkmark & \checkmark & \checkmark & \checkmark\\
10-m to 500 hPa bulk wind difference & \checkmark & \checkmark & \checkmark & \checkmark & \checkmark \\
10-m to 850 hPa bulk wind difference & \checkmark & \checkmark & \checkmark & \checkmark & \checkmark \\

700, 500, and 200 hPa geopotential height & \checkmark & \checkmark & \checkmark & \checkmark & \checkmark \\
850, 500, and 200 hPa zonal wind speed & \checkmark & \checkmark & \checkmark & \checkmark & \checkmark \\
850, 500, and 200 hPa meridional wind speed & \checkmark & \checkmark & \checkmark & \checkmark & \checkmark  \\
850, and 700 hPa temperature & \checkmark & \checkmark & \checkmark & \checkmark & \checkmark \\
850 and 500 hPa specific humidity & \checkmark & \checkmark & \checkmark & \checkmark & \checkmark \\

Surface-based CAPE & & & & & \checkmark \\
Surface-based CIN & & & & & \checkmark \\
0--3 km AGL storm-relative helicity &  & & & & \checkmark\\
6-hourly accumulated precipitation &  & & & & \checkmark \\
2-m AGL specific humidity & & & & & \checkmark \\
Surface pressure & & & & & \checkmark \\

 \hline
\end{tabularx}
\caption{Base predictors used for model training.}
\label{diags-table}
\end{table}

\begin{table}
\centering
\footnotesize
\setlength{\tabcolsep}{4pt} 
\renewcommand{\arraystretch}{0.8} 
\begin{tabular}{lp{0.75\textwidth}}
\toprule
\textbf{Step} & \textbf{Description} \\
\midrule
1 & \textbf{Spatial Interpolation:} Inverse distance weighting interpolation to 80-km grid over CONUS. \\
  & \textit{Details:} Gaussian kernel (25 km std dev), 120 km search radius. \\
\midrule
2 & \textbf{Initialization:} 00 UTC daily. \\
\midrule
3 & \textbf{Lead Times:}  Days 1--8 (6-hourly time step within each range): \\
  & \quad Day 1: 12--36 h, Day 2: 36--60 h, Day 3: 60--84 h, Day 4: 84--108 h, \\
  & \quad Day 5: 108--132 h, Day 6: 132--156 h, Day 7: 156--180 h, Day 8: 180--204 h \\
\midrule
4 & \textbf{Grid Processing:} For each lead time range and 80-km grid point: \\
  & \quad a. Extract variables' value for each time step. \\
  & \quad b. Calculate variables' mean from 5x5 neighborhood grid points for each time step. \\
\midrule
5 & \textbf{Labeling:} Binary label for each grid point and lead time range: \\
  & \quad - 1: Severe event within 40-km radius in 24-hour period. \\
  & \quad - 0: Otherwise. \\
\midrule
6 & \textbf{Normalization:} Calculate \& apply normalization (mean, std dev) using 2018--2023 forecasts. \\
\bottomrule
\end{tabular}
\caption{Dataset Preprocessing Steps}
\label{tab:pseudocode_preprocessing}
\end{table}

The model variables and dataset preprocessing details are provided in Table \ref{diags-table}. Instead of using all available forecast variables, a subset of key variables known to indicate favorable synoptic conditions for severe weather is selected as input. Table \ref{diags-table} presents the forecast variables used in the experiment. Specifically, ``GFS'' contains the same variables as the ``Pangu'' variants, with the different Pangu variants representing different initial conditions. ``GFS+Convect'' includes several convective parameters and surface variables available from the GFS model output which are not available in the Pangu-Weather model output. This allows for a comparison of the importance of convective parameters in severe weather forecasting, especially since most AI weather models lack these parameters.

The preprocessing details are provided in Table \ref{tab:pseudocode_preprocessing}. The gridded forecast output was upscaled onto an 80-km grid covering all of CONUS, similar to the pre-processing applied in \citep{Sobash2020-zs}. Here, we use inverse distance weighting (IDW) to estimate values at 80-km grid points. The IDW weights are determined by a Gaussian kernel with a 25-km standard deviation, applied within a 120-km search radius. For each 80-km grid point across the CONUS, and for daily initialization times at 00 UTC, forecast data is aggregated for eight lead time ranges (days 1--8, 12 UTC -- 12 UTC). For every 6-hourly time step in each 24-hour lead time range, the output variables at each grid point are extracted. Additionally, the mean of these variables over a 5x5 neighborhood centered on the grid point is calculated. Similar to findings in \cite{hill2023}, adding additional statistics such as ``min'' and ``max'' or increasing the neighborhood size showed no significant improvement in model performance.

The distances to severe weather events from the NOAA Storm Prediction Center (SPC) archive were recorded, and a binary label is assigned to each grid point indicating whether a severe report occurred within a 40-km radius during the corresponding 24-hour period. While this distance information was not used as a predictive feature, it was recorded to allow for flexibility in adjusting the proximity threshold for other potential applications. The normalization constants, mean and standard deviation, were calculated using all forecasts from 1 January 2018--31 December 2023. The construction of the preprocessed normalized dataset varies depending on the post-processing framework used, as described in the following Section \ref{sec: post-processing framework}.

\subsection{Post-processing Forecast with Decoder-only Transformer}\label{sec: post-processing framework}
\begin{figure}[t]
 \noindent\includegraphics[width=39pc,angle=0]{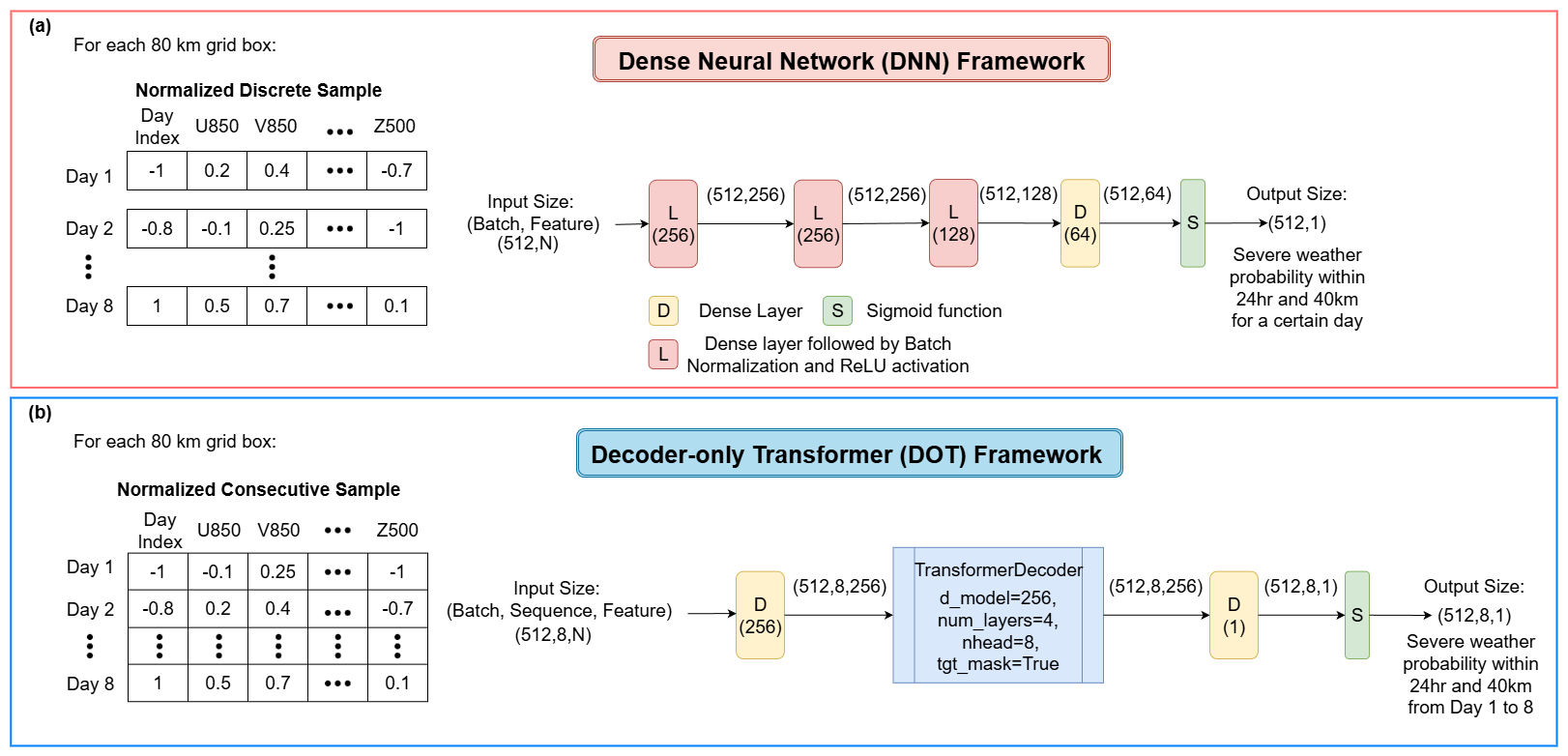}\\
 \caption{Comparison of post-processing frameworks: (a) The Dense Neural Network (DNN) framework takes a discrete sample with an input shape of (Batch, Feature) and outputs a severe weather probability for a single day, with an output shape of (Batch, 1). Shapes above arrows indicate the output shape of each layer. (b) The Decoder-only Transformer (DOT) framework utilizes a consecutive sample of weather forecasts from Day 1 to Day 8, with an input shape of (Batch, Sequence, Feature), and outputs severe weather probabilities for each day from Day 1 to Day 8, resulting in an output shape of (Batch, Sequence, 1). Description of hyperparameters for the PyTorch \citep{paszke2019pytorch} TransformerDecoder (``d\_model'': embedding/layer dimension, ``num\_layers'': number of stacked layers, ``nhead'': number of attention heads, ``tgt\_mask=True'': enables causal masking) are further elaborate in Section \ref{sec: post-processing framework}.}\label{fig: framework}
\end{figure}

Previous work by \cite{Sobash2020-zs} used a dense neural network (DNN) for post-processing, as shown in Fig. \ref{fig: framework}a. The DNN takes a vector of discrete forecast samples and predicts the probability of severe weather events for a given location and lead-time. For each forecast day and grid point, the features include those listed in Table \ref{diags-table} at each 6-hourly output time within the 24-hour forecast period. Instead of using features valid within a specific 24-hour period as input to a DNN, the decoder-only transformer network (DOT; \citealt{vaswani2017attention}, \citealt{radford2019language}) uses forecast information across all forecast days, as shown in Fig. \ref{fig: framework}b. The input is a matrix instead of a vector, with rows representing lead time and columns representing features, and the output is a sequence of severe weather probabilities for each leadtime at a single grid point. This mimics how tokens in large language modeling are structured, where a sentence is broken down into a series of tokens (i.e., words) stacked over the row dimension, and each token has a certain embedding represented in the column dimension.

Compared to the DNN, using the DOT in this way allows the model to use forecast information at multiple days for a given forecast point, and the attention mechanism learns how to relate variables to one another. The specific architecture uses a `TransformerDecoder` with several key hyperparameters detailed in Fig. \ref{fig: framework}b. The `d\_model' parameter defines the dimensionality of the embeddings and internal layers, essentially setting the capacity for representing features at each time step. The model consists of `num\_layers' identical decoder layers stacked sequentially, allowing for hierarchical processing of the input sequence. Within each layer, the multi-head attention mechanism uses `nhead' parallel attention heads, enabling the model to simultaneously focus on different types of relationships between forecast variables across different lead times when predicting the severe weather probability. Additionally, the attention mechanism incorporates a crucial look-ahead mask, enabled by `tgt\_mask=True'. This mask enforces the auto-regressive property and is used both in training and inference, ensuring that when predicting the severe weather probability for a specific lead time (e.g., Day 4), the model can only attend to the forecast information from the current and preceding lead times (Days 1-4) and is prevented from accessing information from future lead times (Days 5-8). This auto-regressive property is enforced to ensure the model learns the temporal evolution of atmospheric features leading up to an event, preventing reliance on forecast data from future lead times which could obscure causal relationships. This is analogous to sentence generation where the model cannot look ahead at future tokens when predicting the current token. During inference, we maintain `tgt\_mask=True'; the model receives the consecutive forecast sample from Day 1 to Day 8 as input and simultaneously outputs the severe weather probability for each day in the sequence (Day 1 to Day 8), from which we extract the desired probability for each day.

\subsection{Training and Verification}\label{subsection: training and verification}

\subsubsection{Model and Training Setup}\label{subsubsection: model and training setup}
The DNN framework is built from a series of dense layers (Fig. \ref{fig: framework}a). It begins with three layers, denoted as 'L', which represent a linear layer followed by Batch Normalization and ReLU activation, with output shapes sequentially transforming from (B, N) to (B, 256), then to (B, 256), and finally to (B, 128), where B is the batch size. This is followed by two dense layers, with output shapes changing from (B, 128) to (B, 64) and then to (B, 1). The final output of the dense layer is passed through a sigmoid function to produce the probability output. In contrast, the DOT framework begins with a dense layer that projects the input of shape (B, 8, N) to a higher-dimensional representation of shape (B, 8, 256), where N is the number of features (Fig. \ref{fig: framework}b). This representation is then fed into a PyTorch's ``TransformerDecoder'' block, configured with parameters d\_model=256, num\_layers=4, nhead=8, and tgt\_mask=True. The TransformerDecoder processes this input and maintains the shape as (B, 8, 256). Finally, a second dense layer reduces the output dimensionality to (B, 8, 1), producing the sequence of severe weather probabilities. 

To evaluate the effectiveness of the DOT model, we applied it to all forecast datasets listed in Table \ref{diags-table}. The DNN was implemented for the ``GFS" and ``GFS+Convect''  model. The DNN serves as a baseline to quantify the performance gains achieved by the transformer architecture. Furthermore, the DOT(GFS) model is included to provide a direct comparison with Pangu-Weather with an identical set of features. This allows us to assess how GFS performs when excluding convective parameters, offering insights relevant to models like Pangu-Weather that also do not explicitly incorporate these parameters. Finally, the mean of the ensemble forecast is generated by taking the mean probability forecast from those three individually trained models which are DOT(Pangu-GFS), DOT(Pangu-HRES) and DOT(GFS+Convect).


Each model was trained on forecast data from January 1, 2018, to December 31, 2023. The validation set excluded the period between May 15, 2023, and August 15, 2023. The test period forecasts were initialized using data from February 1, 2024, to July 22, 2024, and with forecast valid times extending to July 31, 2024. This short period exclusion in the validation set was a minor adjustment to maximize the training dataset, as validation was primarily used for early stopping. The training objective is identical to that of a binary classification task, which uses binary cross-entropy loss. A batch size of 512 was consistently used for both the transformer and DNN frameworks. Both frameworks utilized a Cosine Annealing Warm Restarts scheduler. This scheduler cyclically adjusts the learning rate, starting with an initial period of 10 epochs, and subsequently maintaining the period between restarts after each cycle, with a minimum learning rate of 1e-7. The optimizer for both was Stochastic Gradient Descent with a momentum of 0.9 and a weight decay of 1e-4. The initial learning rate for the transformer was set to 6e-4, while for the dense neural network it was 8e-4. Training was halted if no improvement in the validation Brier Skill Score was observed for 20 consecutive epochs. Validation BSS typically plateaued after approximately 150 epochs for the transformer model and around 50 epochs for the dense neural network model. Finally, the training was performed on a single Nvidia L40 GPU. Additional ablation tests regarding ``d\_model'' and ``num\_layers'' are provided in Fig. S1 and show that there was no significant impact on performance when the other settings remained the same.

\subsubsection{Verification Metrics}
To evaluate the performance of the severe weather probability forecasts, we employ a suite of verification metrics including the Brier Skill Score (BSS; \cite{wilks2011statistical}), Precision-Recall (PR), and Receiver Operating Characteristic (ROC) curves. 

The Brier Score (BS) quantifies the mean squared error of probabilistic forecasts, calculated as:
\begin{equation}
    \text{Brier score} = \frac{1}{N} \sum_{t=1}^{N} (f_t - o_t)^2
    \label{eq:brier_score}
\end{equation}
where \(N\) is the total number of forecasts, \(f_t\) is the forecast probability for the \(t^{th}\) event, and \(o_t\) is the binary outcome (0 or 1) of the \(t^{th}\) event. To assess the skill relative to a reference forecast, we utilize the BSS, defined as:
\begin{equation}
    \text{Brier skill score} = 1 - \frac{BS_f}{BS_{ref}}
    \label{eq:brier_skill_score}
\end{equation}
where \(BS_f\) is the Brier Score of the forecast model and \(BS_{ref}\) is the Brier Score of a reference forecast. Here, the reference forecast \(BS_{ref}\) is the Brier Score of a simple climatological forecast, where the forecast probability is the static, overall frequency of severe weather events (the base rate) in the provided test period dataset or period. Positive BSS values indicate improvement over the reference, with 1 representing perfect skill, 0 indicating no skill difference from the reference, and negative values indicating worse performance than the reference.  

Beyond scalar metrics, we also employ the PR curve and ROC curve to evaluate forecast discrimination across varying probability thresholds. In both PR and ROC contexts, \textit{True Positives (TP)} are correctly predicted severe weather events, \textit{False Positives (FP)} are incorrectly predicted severe weather events, \textit{False Negatives (FN)} are missed severe weather events, and \textit{True Negatives (TN)} are correctly predicted non-severe weather events. Precision, defined as \( \frac{TP}{TP+FP} \), represents the proportion of correctly predicted severe weather events out of all predicted events, while Recall (i.e, the True Positive Rate, TPR, or the probability of detection, POD), defined as \( \frac{TP}{TP+FN} \), represents the proportion of correctly predicted severe weather events out of all actual events.  The PR curve visualizes the trade-off between precision and recall as the probability threshold for classifying an event as severe is varied. A large area under the PR curve (PRAUC) signifies strong performance with both high precision and high recall. Similarly, the ROC curve illustrates the trade-off between the True Positive Rate (TPR) and the False Positive Rate (FPR), defined as \( \frac{FP}{FP+TN} \), as the probability threshold changes. A large area under the ROC curve (ROCAUC) indicates good discrimination, reflecting high TPR and low FPR across different thresholds.

\subsubsection{Explainable AI with Integrated Gradients}
To gain insights into the decision-making process of our decoder-only transformer framework, we employ Explainable Artificial Intelligence (XAI) techniques, specifically Integrated Gradients (IG; \cite{sundararajan2017axiomatic}). IG is a feature attribution method that aims to explain the relationship between input features and model predictions by accumulating gradients along a linear path from a baseline input (the matrix of zeros represents the forecast mean state) to the actual input. In our application, we customize IG to diagnose feature sensitivity for specific forecast lead times, focusing on Day 3, 4, 5, and 6. For each of these days and grid points, we first filter test samples based on whether the models' predicted severe weather probability from DOT(GFS+Convect), DOT(GFS), DOT(Pangu-GFS) and DOT(Pangu-HRES) all exceed a predefined threshold. The predefined threshold is based on the calibration curve shown in Fig. \ref{fig: calibration_curve} with thresholds exceeding 0.3, 0.25, 0.2, and 0.15 for forecast Days 3, 4, 5, and 6 respectively. Then, for the samples that surpass this threshold for a given day, we compute the IG to attribute the prediction back to the input features. The attributions are aggregated across the 6-hourly time steps within each 24-hour forecast period, providing daily-scale feature importance rather than subdaily temporal resolution. 
By aggregating these attributions across multiple samples for each target day, we can identify the key input forecast features that are most influential in the model's prediction of severe weather events at different forecast horizons, thus providing interpretability to the transformer's complex decision process.

\section{Forecast Verification Period Overview}
\begin{figure}[t]
\centering
 \noindent\includegraphics[width=13cm]{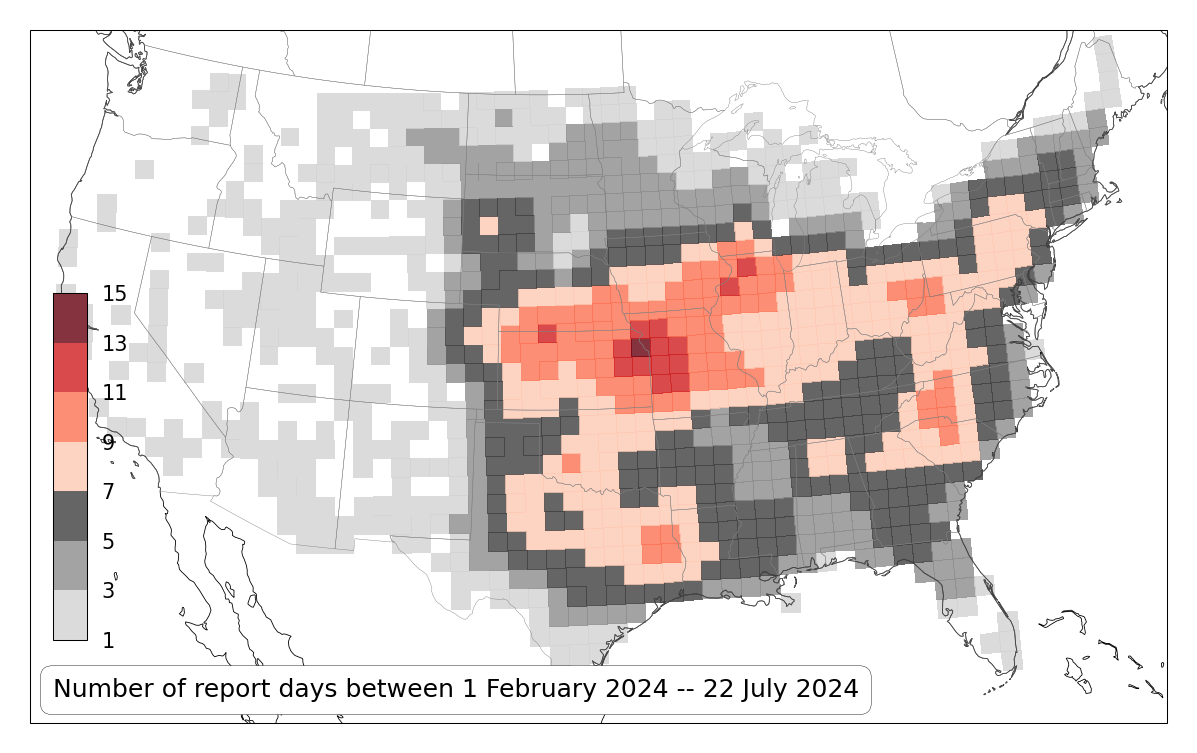}\\
 \caption{Number of days (12 UTC -- 12 UTC) with $\geq$ 1 storm report occurring within each 80-km grid box during the verification period (1 February 2024 -- 22 July 2024). Report day totals were smoothed using a Gaussian smoother with $\sigma$ = 80 km. }\label{fig:report-total}
\end{figure}

The forecast verification period used in this work (1 February 2024 -- 22 July 2024) was a particularly active period for severe convective weather events within the CONUS. By the end of July 2024, the total number of tornado and wind reports in 2024 received at the Storm Prediction Center was in the top 3 among the last 15 severe weather seasons, while the number of hail reports was slightly below normal \citep{SPCwcm}. Severe weather events occurring during April and May 2024 were particularly frequent and intense. During this two month period, the yearly total of tornado reports went from slightly below average in early April to near the 90th percentile of report totals over the 25 year period between 1999--2023 by the end of May \citep{SPCwcm}. This two-month period included 11 days with $\geq$ 25 preliminary tornado reports across the CONUS and at least a slight risk of severe weather was issued every day somewhere over CONUS by the SPC, with 16/31 days having an enhanced risk, 5 days having a moderate risk, and 1 having a high risk. During the verification period, severe weather events were most common across the central Great Plains into the Midwest (i.e., eastern Colorado through Missouri and Indiana), with $\approx$10\% of days having $\geq$ 1 severe weather report across eastern Kansas and western Missouri (Fig. \ref{fig:report-total}).

The events during Spring 2024 were often strongly forced by large-scale weather systems and thus more predictable compared to severe weather events driven by more subtle small-scale features. The mean large-scale flow pattern over CONUS during May 2024 was characterized by a trough in the western U.S., that was slightly anomalous for the period, a pattern conducive to severe weather across the central U.S. (not shown). The enhanced predictability during the Spring 2024 period provides an opportunity to utilize AI NWP-based convective hazard forecasts, given that these data-driven models can provide more accurate large-scale predictions into the medium-range than traditional NWP approaches; it is hypothesized that less benefit would exist compared to traditional NWP for events driven by storm- and meso-scale processes.

\section{Post-processing Forecast Verification}

\subsection{Forecast Skill versus Lead Time}

\subsubsection{Transformer (DOT) versus DNN Architecture}
\begin{figure}[h!]
    \centering
    \noindent\includegraphics[width=38pc,angle=0]{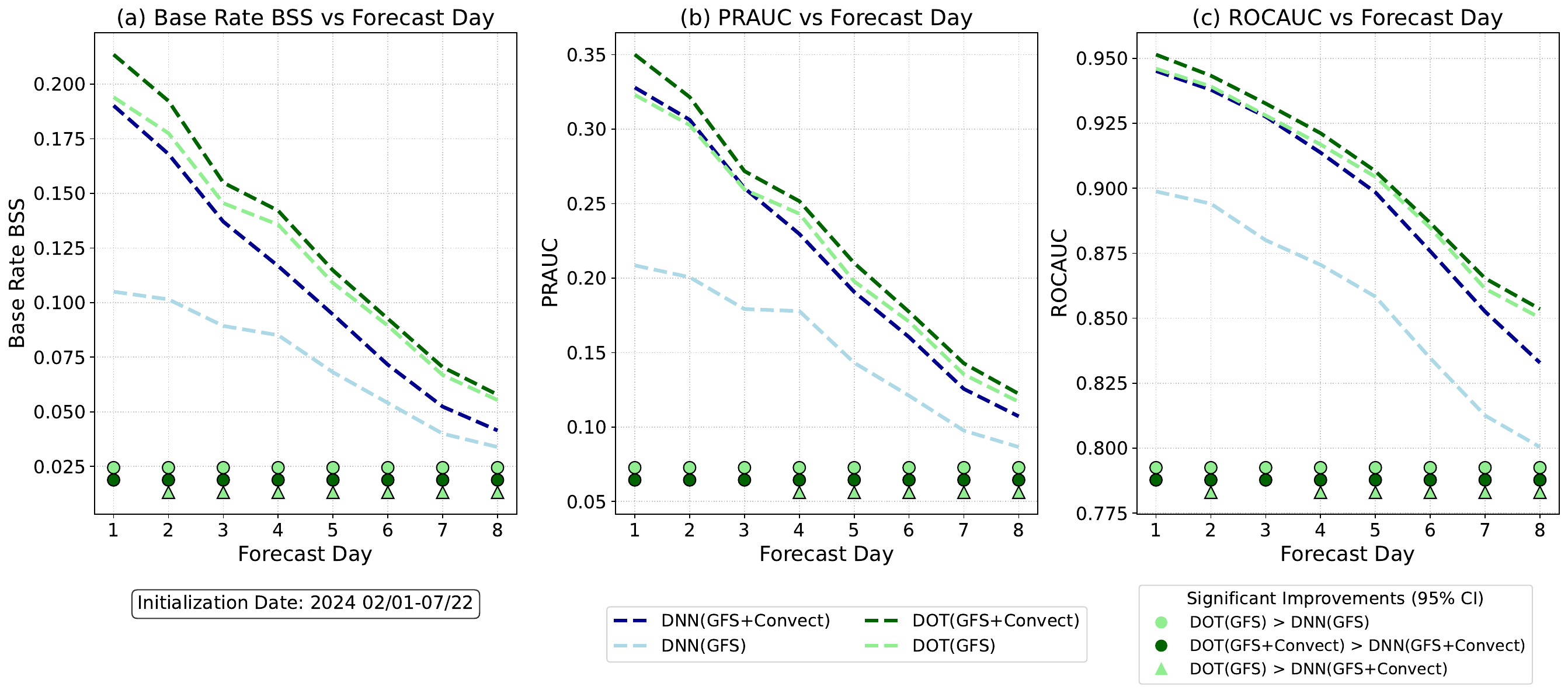}\\
    \caption{The plots show (a) Base Rate BSS, (b) PRAUC, and (c) ROCAUC performance comparison as a function of forecast day between DOT and DNN architectures using GFS forecast. The lines in the bottom right legend represent different models or configuration. Statistical significance at the 95\% confidence level was assessed for the specific model pairs listed in the legend below the plot. For each pair, 1,000 bootstrap resamples of the skill difference were computed at each lead time. A symbol is shown if the corresponding bootstrapped 95\% confidence interval lies entirely above zero, indicating a statistically significant improvement.}
    \label{fig:dot_vs_dnn_gfs}
\end{figure}

A direct comparison between the Decoder-only Transformer (DOT) and the Dense Neural Network (DNN) architectures, both utilizing GFS inputs, reveals the significant advantages of the transformer-based approach (Fig. \ref{fig:dot_vs_dnn_gfs}). The DOT(GFS+Convect) and DOT(GFS) models consistently outperform their DNN counterparts (DNN(GFS+Convect) and DNN(GFS)) across all forecast lead times and performance metrics. Notably, there is a significant performance drop for the DNN model when convective parameters are removed, as seen when comparing DNN(GFS) to DNN(GFS+Convect). In contrast, the DOT(GFS) model, even without explicit convective parameters, significantly outperforms both DNN models. For lead times less than 4 days, DOT(GFS) nearly doubles the BSS performance relative to DNN(GFS) and also significantly outperforms DNN(GFS+Convect) after Day 1. This statistically significant improvement highlights the effectiveness of the transformer architecture in capturing temporal dependencies within the forecast sequence.

\subsubsection{AI NWP (Pangu-Weather) versus NWP (GFS) Forecast}
\begin{figure}[h!]
    \centering
    \noindent\includegraphics[width=38pc,angle=0]{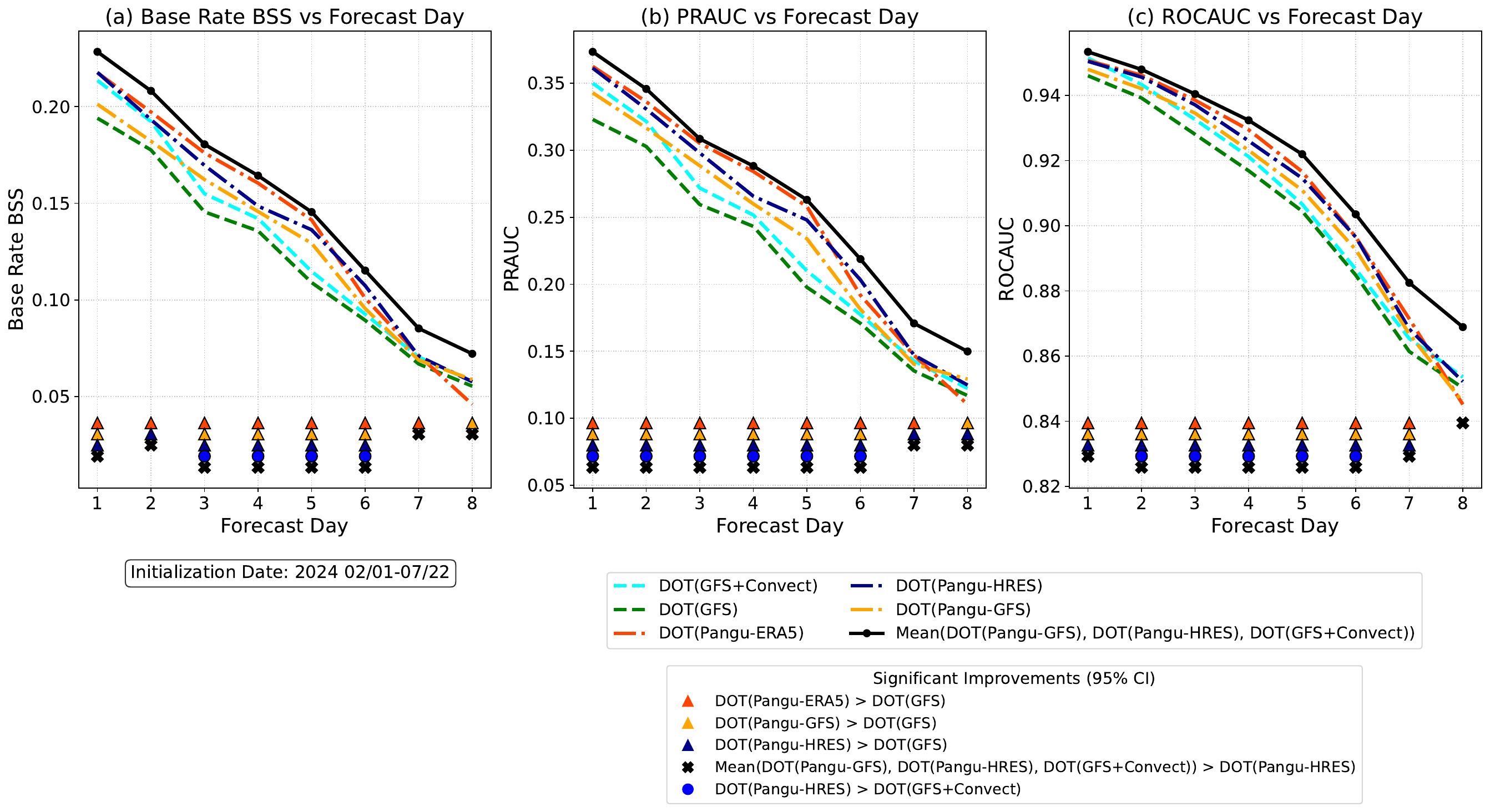}
    \caption{Similar to Fig. \ref{fig:dot_vs_dnn_gfs}, this figure compares the performance of the DOT post-processing architecture, but using Pangu-Weather forecasts versus GFS forecasts as input.}
    \label{fig:pangu_vs_gfs_dot}
\end{figure}

When comparing the performance of the DOT architecture with different forecast sources---AI-driven Pangu-Weather forecasts versus traditional NWP GFS forecasts---the results demonstrate the potential capability of the AI-based forecast, particularly in the medium-range (Fig. \ref{fig:pangu_vs_gfs_dot}). The DOT(Pangu-GFS) model significantly outperforms the DOT(GFS) model in almost all lead times, with the most substantial improvements observed between Days 3 and 6 across all three metrics. Lastly, DOT(Pangu-HRES) significantly outperforms DOT(GFS) in all metrics at nearly all lead times except Day 8 and also outperforms DOT(GFS+Convect), across all three skill metrics from Day 3 to Day 6. DOT(Pangu-HRES) achieves a performance similar to that of DOT(Pangu-ERA5) after Day 5 and surpasses it by Day 8. These findings underscore the value of using AI NWP emulators like Pangu-Weather for generating more skillful medium-range severe weather forecasts.

Finally, the operational model mean (Mean(DOT(Pangu-GFS), DOT(Pangu-HRES), DOT(GFS+Convect))), representing the ensemble average output probability of `DOT(GFS+Convect)', `DOT(Pangu-GFS)', and `DOT(Pangu-HRES)', consistently attains the best skill across all metrics and forecast days. This highlights the power of ensemble forecasting with potentially well-calibrated models, where averaging can effectively mitigate individual model errors and yield overall superior and more robust predictions. Our analysis in the following sections will focus solely on the verification of Day 3 to Day 6 forecasts where the differences between the GFS and Pangu-Weather forecasts are largest and statistically significant.

\subsection{Sensitivity of Probability Threshold versus Lead Time}

\begin{figure}[t]
 \noindent\includegraphics[width=38pc,angle=0]{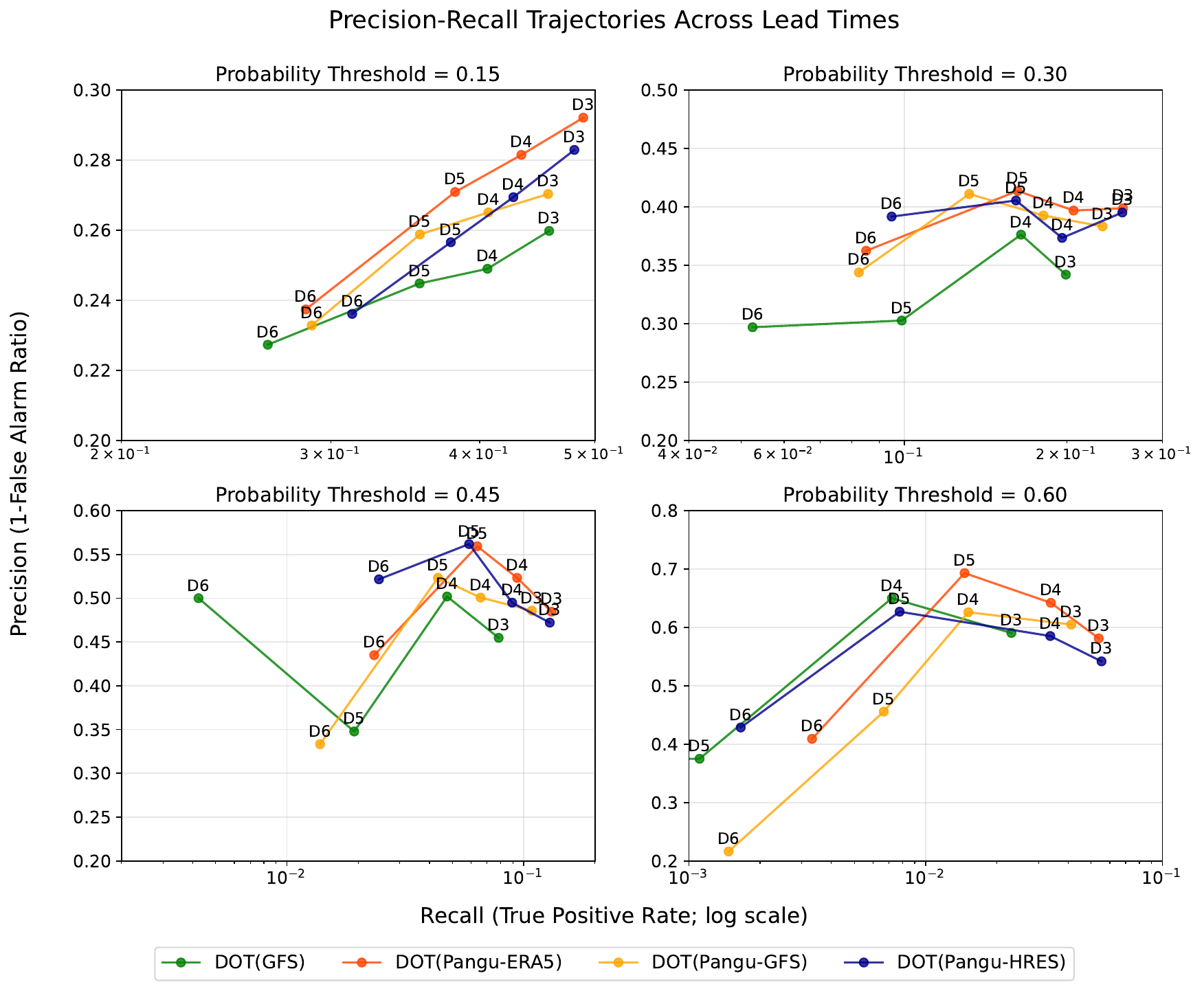}\\
 \caption{Precision-Recall curves for different forecast models (DOT(GFS), DOT(Pangu-ERA5), DOT(Pangu-GFS), DOT(Pangu-HRES)) across lead times from Day 3 (D3) to Day 6 (D6). Each panel displays the performance at a fixed probability threshold: (a) 0.15, (b) 0.30, (c) 0.45, and (d) 0.60. Precision (1 - False Alarm Ratio) is plotted against the Recall (True Positive Rate) on a logarithmic scale.}\label{fig: pr_trajectories_multi_panel}
\end{figure}

Fig. \ref{fig: pr_trajectories_multi_panel} illustrates the performance evolution of different forecast models by plotting Precision-Recall (PR) trajectories across lead times from Day 3 to Day 6. It highlights how model skill degrades with increasing forecast lead time at fixed probability thresholds. Generally, for most models and thresholds, the points corresponding to later lead times (e.g., Day 5, Day 6) are located towards the lower-left relative to earlier lead times (e.g., Day 3, Day 4), indicating a decrease in both precision and recall.

Comparing the models, DOT(Pangu-ERA5) often shows strong performance, particularly at shorter lead times (Day 3, Day 4) and higher thresholds (Fig. \ref{fig: pr_trajectories_multi_panel}c,d), maintaining relatively high precision and recall. DOT(Pangu-HRES) also demonstrates competitive performance, often closely following or slightly underperforming DOT(Pangu-ERA5) at shorter lead times but sometimes showing better retention of skill at Day 5 or Day 6 depending on the threshold. DOT(Pangu-GFS) generally exhibits lower performance than the Pangu-ERA5 and Pangu-HRES variants, especially at longer lead times. The baseline DOT(GFS) consistently shows the lowest performance among the models across most thresholds and lead times.

The choice of probability threshold significantly impacts the observed performance and the relative ranking of models. At lower thresholds (e.g., 0.15, Fig. \ref{fig: pr_trajectories_multi_panel}a), recall is generally higher but precision is lower for all models. As the threshold increases (Fig. \ref{fig: pr_trajectories_multi_panel}a-d), precision improves while recall decreases, and the performance separation between models often becomes more distinct. The specific trajectory shapes reveal nuances; for example, some models might maintain precision better than recall as lead time increases, or vice-versa, depending on the threshold. This type of analysis is crucial for understanding how models behave under different operational decision-making scenarios.

\subsection{Reliability of Forecast Probability}

\begin{figure}[t]
 \noindent\includegraphics[width=37pc,angle=0]{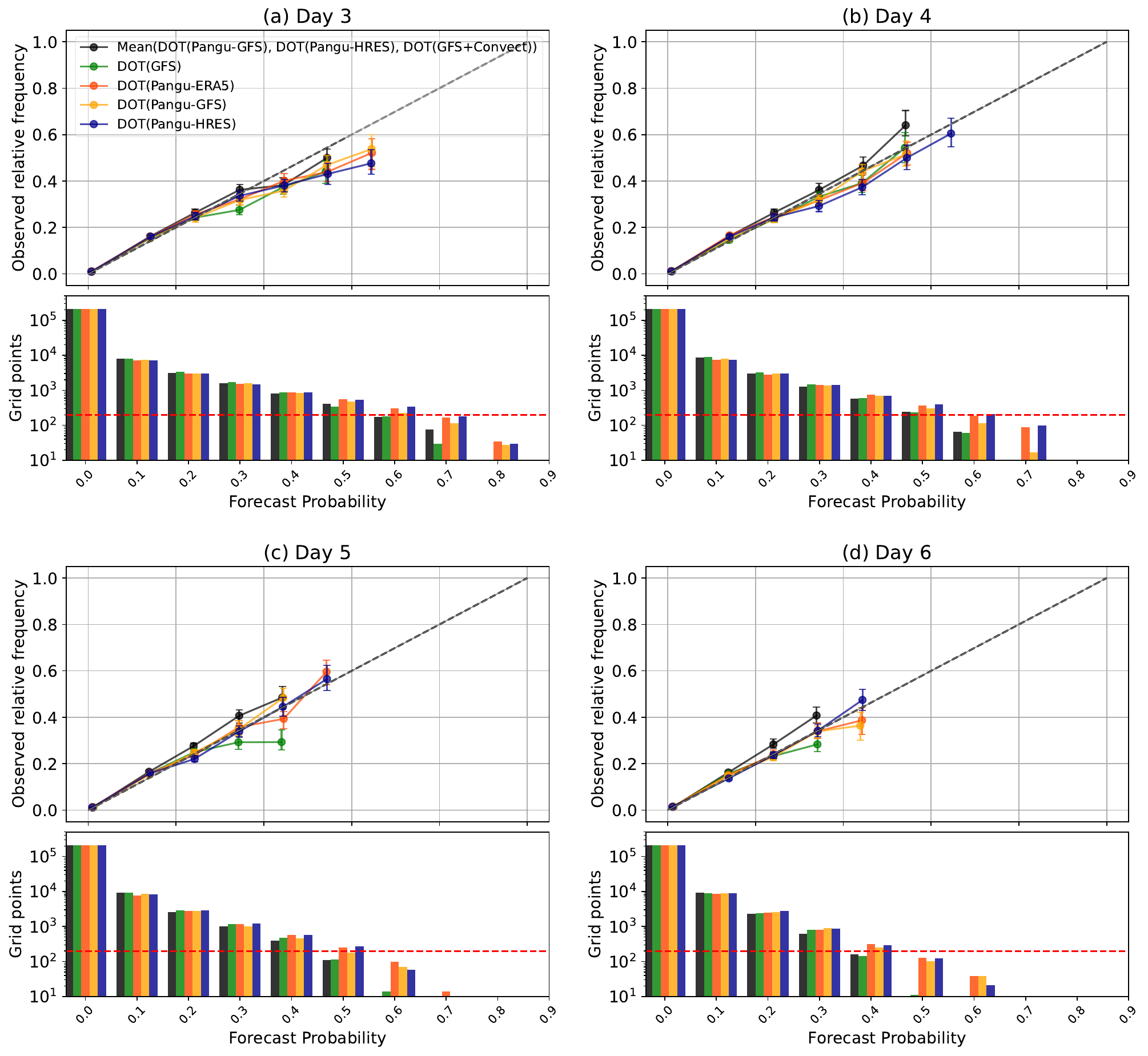}\\
 \caption{Reliability diagrams for severe weather probability forecasts at lead times of (a) Day 3, (b) Day 4, (c) Day 5, and (d) Day 6.  Each panel shows two subplots. The top subplot displays reliability curves for various models. The dashed diagonal line represents perfect reliability (forecast probability equals observed relative frequency). The bottom subplot shows the distribution of forecast probabilities (number of grid points per probability bin), with a logarithmic y-axis to highlight the skew towards lower probabilities. Error bars indicate bootstrapped 95\% confidence intervals. The red dashed horizontal line in the lower subplot indicates a cut-off of 200 samples, below which the reliability curve calculation is not performed.}\label{fig: calibration_curve}
\end{figure}

To assess the calibration of the forecast probabilities, Figure \ref{fig: calibration_curve} presents reliability diagrams for forecast days 3 -- 6. A perfectly calibrated model would exhibit a one-to-one correspondence between the forecast probability and the observed relative frequency of severe weather events.  In such a case, the reliability curve would align perfectly with the diagonal dashed line (representing perfect reliability). Points above this diagonal indicate under-forecasting (the model's predicted probabilities are too low), while points below the diagonal indicate over-forecasting (the model's predicted probabilities are too high).

As the forecast lead time increases from Day 3 to Day 6, two trends are apparent. First, the maximum forecast probability assigned by the models decreases. This is because as the lead time increases, the model is less confident with the higher probability, and also the models become less confident in their predictions, reflecting the increasing uncertainty associated with longer-range forecasts. Second, the reliability curves for all individual models tend to exhibit slight over-forecasting at Day 3, particularly for observed relative frequencies greater than 30\%, but this diminishes somewhat at longer lead times. Interestingly, the DOT(GFS) model tends to struggle with predicting probability greater than 50\% compared to other Pangu models.

The Mean(DOT(Pangu-GFS), DOT(Pangu-HRES), DOT(GFS+Convect)) shows a different behavior. While it generally performs the best (as indicated by its superior Brier Skill Score in Fig. \ref{fig:pangu_vs_gfs_dot}), it exhibits a tendency to under-forecast at higher probability intervals. This under-confidence is not ideal for perfect calibration, it can be beneficial for the overall Brier Skill Score (BSS). Due to the Brier Score's quadratic nature, it penalizes large errors disproportionately. For example, an incorrect forecast of 90\% receives more than double the penalty of an incorrect 60\% forecast. By systematically avoiding high-confidence predictions, the model mitigates the risk of these large, heavily penalized errors, which often improves the average score. In summary, while exhibiting some deviations from perfect calibration, all models produce generally reliable forecasts, and the operational model mean still produces the best BSS.

\subsection{Forecast Skill Time Series}

\begin{figure}[t]
 \noindent\includegraphics[width=34pc,angle=0]{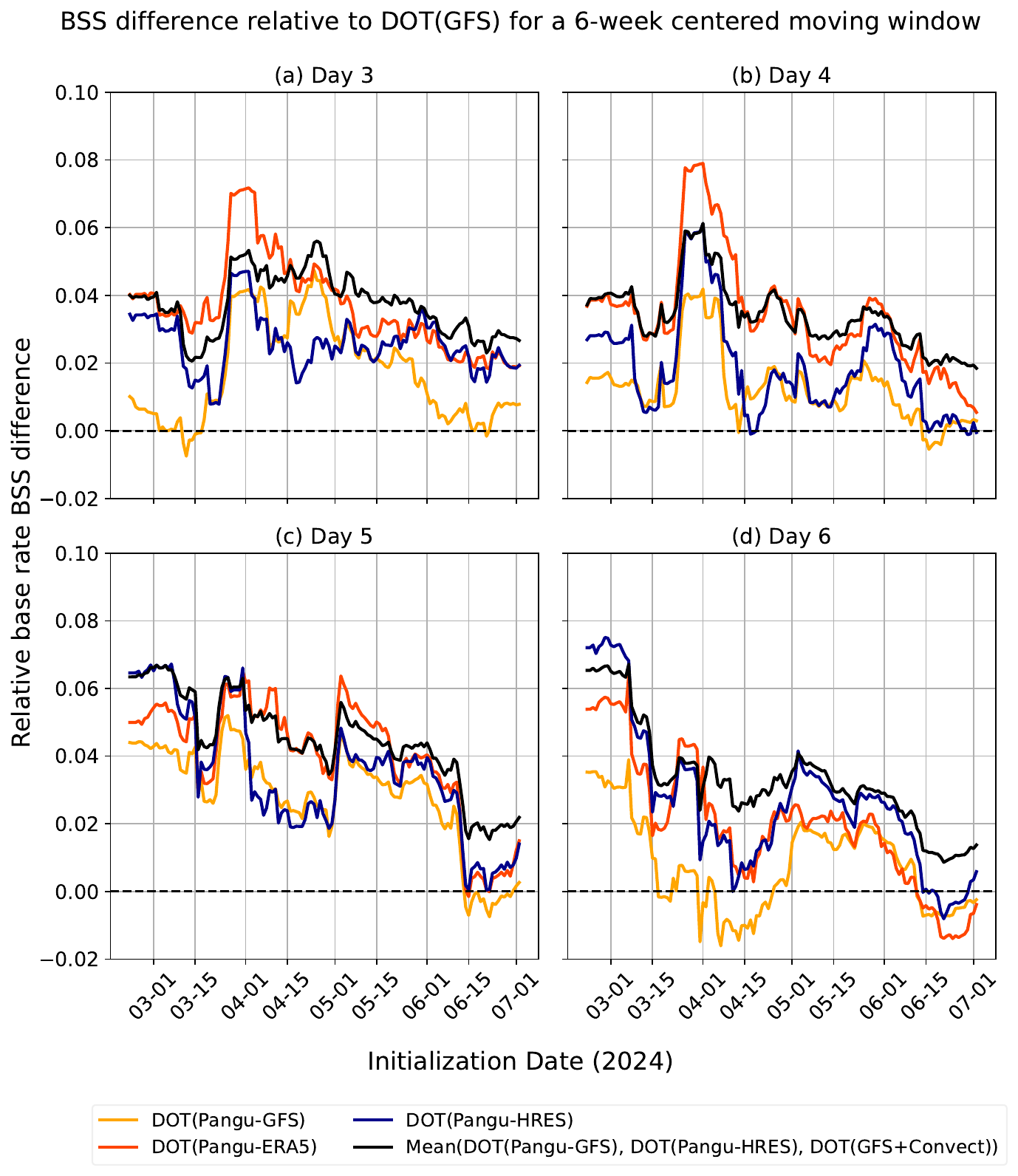}\\
 \caption{Time series of the Brier Skill Score (BSS) difference relative to DOT(GFS) for severe weather probability forecasts at lead times of (a) Day 3, (b) Day 4, (c) Day 5, and (d) Day 6. The BSS difference is calculated using a 6-week centered moving window applied to the samples, reducing sensitivity to daily forecast fluctuations. The x-axis represents the centered initialization time of the forecasts within the moving window. The y-axis represents the BSS difference, where positive values indicate improved skill compared to DOT(GFS).}\label{fig: temporal_bss}
\end{figure}

Here we investigate the temporal evolution of the BSS difference relative to the DOT(GFS) model for Day 3 -- 6 (Fig. \ref{fig: temporal_bss}a-d, respectively), calculated using a 6-week centered moving window. Several interesting patterns emerge from this time series analysis. First, across all lead times, most models demonstrate a positive BSS difference relative to DOT(GFS) for the majority of days, indicating improved skill that is persistent through the evaluation period. However, the magnitude of this improvement varies depending on models and lead time. For instance, DOT(Pangu-GFS) performs consistently worse than DOT(GFS) for a stretch of days in late March and early April 2024 at Day 6, and for shorter stretches at earlier lead-times. A case during this period will be examined in a later section. Secondly, there is a noticeable temporal variation in the BSS difference. For instance, the BSS difference for many models tends to peak in late March and early April before gradually decreasing into the summer months. This fluctuation could be attributed to stronger synoptic forcing and more organized weather systems during the spring transition period, which may be better captured by the Pangu-Weather compared to GFS. Furthermore, the Pangu models exhibit considerable fluctuation in their BSS difference throughout the period, suggesting a sensitivity to specific forecast initialization times and potentially varying performance across different weather regimes. Periods of higher BSS difference might correspond to more active storm seasons or periods with stronger, more predictable synoptic-scale forcing. Conversely, periods of lower BSS difference or even negative differences could indicate challenges in specific weather situations or regimes where the strengths of the Pangu-Weather are less pronounced, or even where DOT(GFS) performs well. The Mean(DOT(Pangu-GFS), DOT(Pangu-HRES), DOT(GFS+Convect)), while also exhibiting temporal variability, generally maintains a more stable and consistently positive BSS difference, suggesting a degree of robustness achieved through ensemble averaging that mitigates some of the fluctuations seen in individual models. Similar findings are also supported by the data in Fig. S2, where DOT(GFS+Convect) is used as the baseline.

\subsection{Forecast Skill's Dependence on Forecasted Convective Parameters}

\begin{figure}[t]
 \noindent\includegraphics[width=38pc,angle=0]{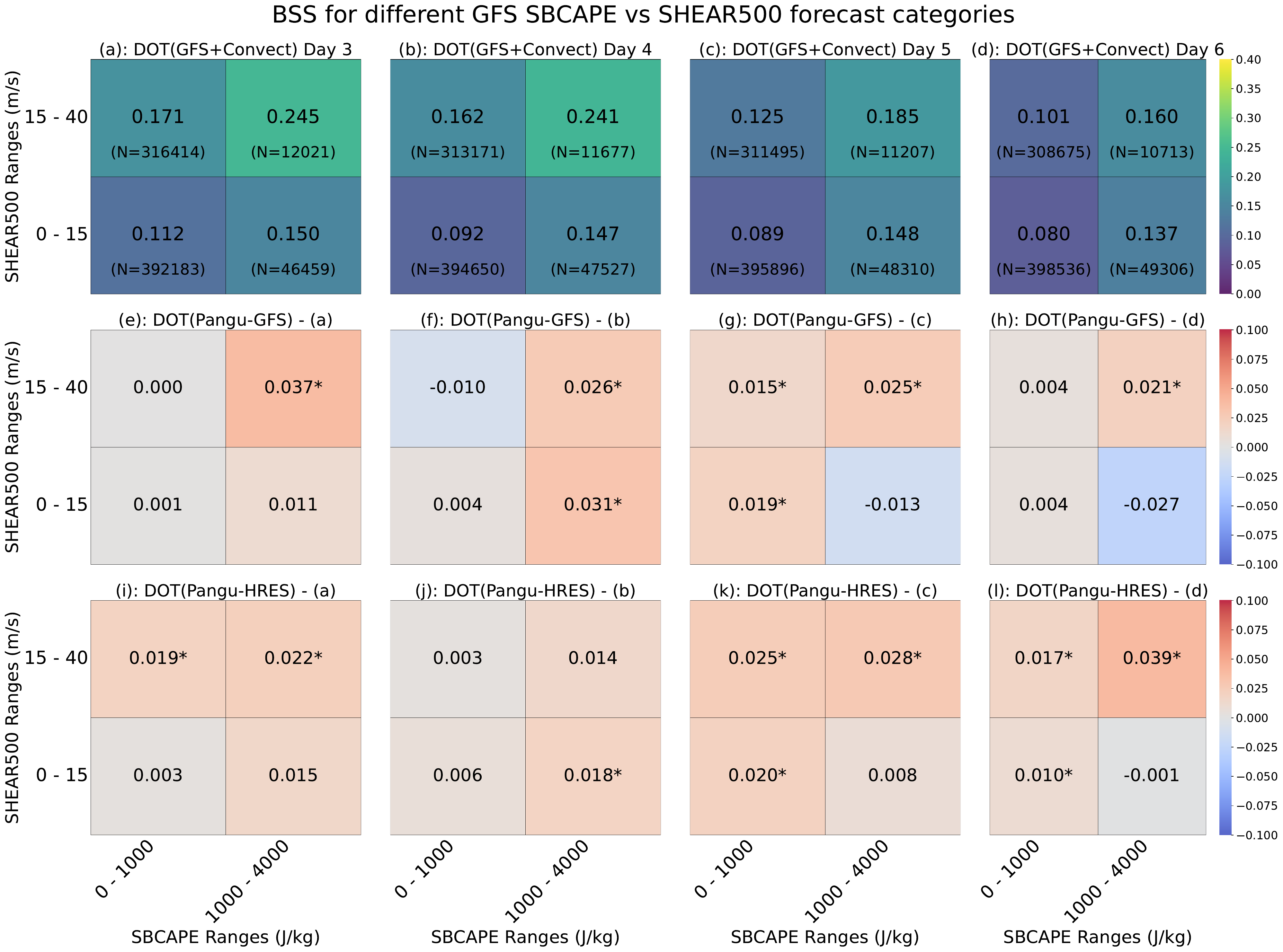}\\
 \caption{Brier Skill Score (BSS) and relative BSS difference conditioned on GFS-forecasted convective parameters. Panels (a-d) show the BSS of the DOT(GFS+Convect) model for Day 3 to 6, respectively, stratified by Surface-Based CAPE (SBCAPE) and surface to 500mb wind shear (SHEAR500) ranges. Panels (e-h) show the BSS difference between DOT(Pangu-GFS) and DOT(GFS+Convect) (DOT(Pangu-GFS) $-$ DOT(GFS+Convect)) for the same days and convective parameter ranges. Panels (i-l) show the BSS difference between DOT(Pangu-HRES) and DOT(GFS+Convect) (DOT(Pangu-HRES) $-$ DOT(GFS+Convect)). Positive values (red) indicate improved skill of Pangu models relative to GFS, while negative values (blue) indicate reduced skill. The number of forecast samples (N) within each SBCAPE/SHEAR500 bin is indicated in parentheses for the GFS model (panels a-d). Pair-wise significance testing similar to Fig. \ref{fig:dot_vs_dnn_gfs} to compare against DOT(Pangu-GFS) and DOT(Pangu-HRES) with DOT(GFS+Convect) for different forecasted convective parameters regimes. With asterisks mark after score indicates statistically significantly better than DOT(GFS+Convect). }\label{fig: forecast_env_verif}
\end{figure}

Figure \ref{fig: forecast_env_verif} examines the BSS of severe weather forecasts conditioned on GFS-forecasted SBCAPE and surface to 500mb wind speed difference (SHEAR500) averaged over each lead time interval for Day 3 to 6. This allows us to see if the differences between the operational and AI models are a function of environmental characteristics. The first row (Fig. \ref{fig: forecast_env_verif}a-d) displays the BSS of the DOT(GFS+Convect) model across different ranges of SBCAPE and SHEAR500. Generally, for DOT(GFS+Convect), the BSS tends to increase with both increasing SBCAPE and SHEAR500, especially moving towards the highest SBCAPE range and higher shear categories. This indicates that GFS exhibits better skill in strongly-forced synoptic regimes, where its performance is less reliant on the intricacies of its convective parameterization scheme.

The subsequent rows illustrate the BSS difference relative to DOT(GFS+Convect) for DOT(Pangu-GFS) (Fig.~\ref{fig: forecast_env_verif}e-h) and DOT(Pangu-HRES) (Fig.~\ref{fig: forecast_env_verif}i-l). DOT(Pangu-GFS) consistently shows positive BSS differences across most SBCAPE and SHEAR500 ranges for Day 3 and Day 4, indicating improved skill over DOT(GFS+Convect), particularly in the higher shear and SBCAPE categories. For Day 5 and Day 6, the advantage of DOT(Pangu-GFS) diminishes, and in some high SBCAPE and low shear regimes, DOT(GFS+Convect) even shows better performance. DOT(Pangu-HRES) exhibits a more pronounced and consistent skill improvement over GFS across all lead times and convective parameter ranges. Notably, DOT(Pangu-HRES) demonstrates particularly strong positive BSS differences in the higher SBCAPE and higher SHEAR500 regimes for lead times greater than 4 days, suggesting that it is especially adept at enhancing severe weather forecasts in strongly forced convective environments.

Surprisingly, both DOT(Pangu-GFS) and DOT(Pangu-HRES), despite not directly utilizing SBCAPE as input features in their native forecast products, still demonstrate a clear ability to improve upon GFS forecasts, particularly in convective environments characterized by high SBCAPE and strong shear. This suggests that the forecast skill advantage inherent in the Pangu models, when coupled with our advanced post-processing framework, enables the system to implicitly capture and leverage information related to convective instability from other input variables. Even without explicit convective parameters as direct inputs, this highlights the powerful pattern recognition capabilities of AI models, especially when enhanced by sophisticated post-processing, in forecasting complex meteorological phenomena.

\subsection{Feature Sensitivity}
\begin{figure}[t]
 \centering\noindent\includegraphics[width=38pc,angle=0]{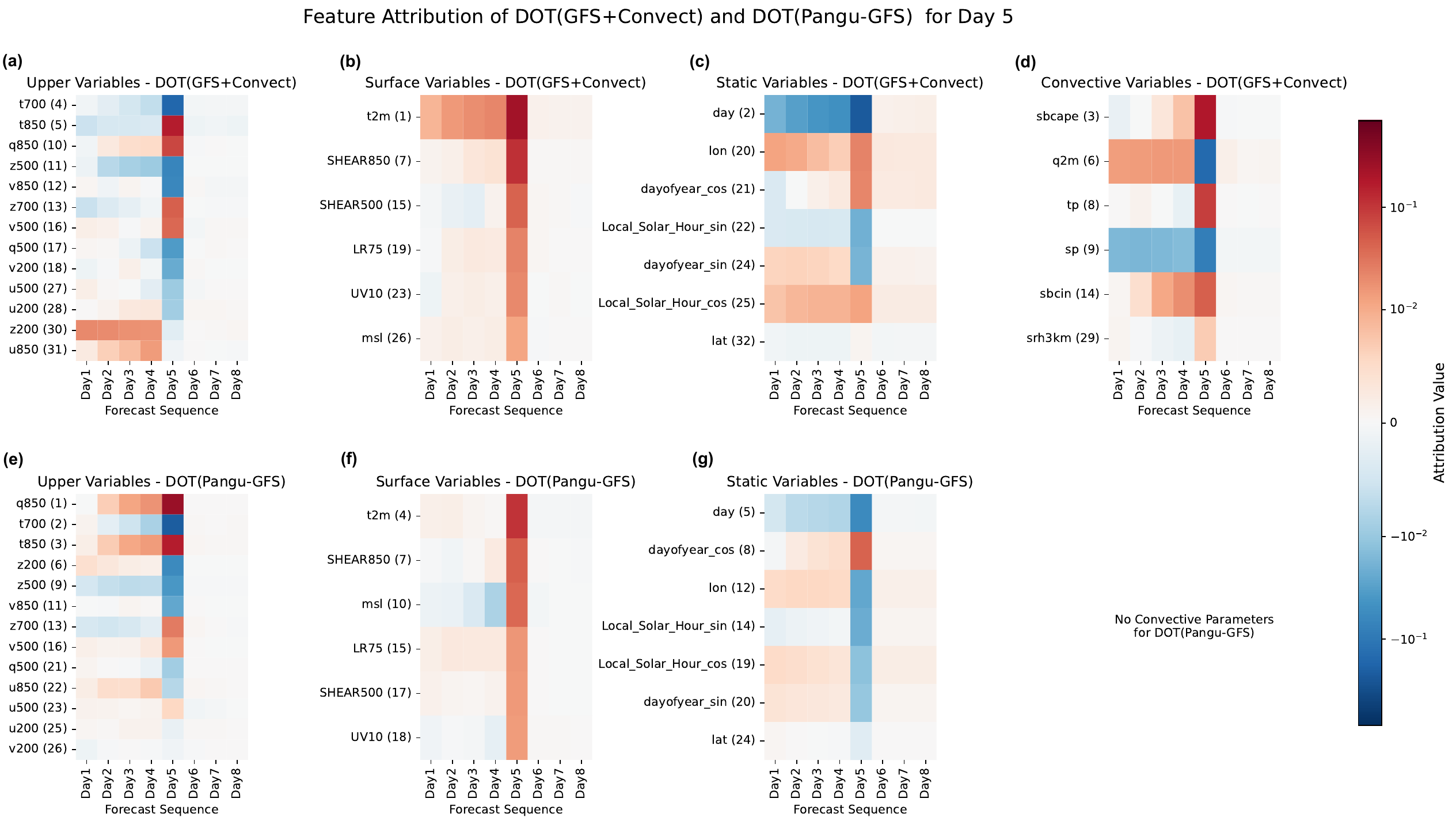}\\
 \caption{Integrated Gradients (IG) feature attribution heatmaps for Day 5 severe weather forecasts. Panels (a)-(d) show results for DOT(GFS+Convect), and panels (e)-(g) show results for DOT(Pangu-GFS). Each heatmap displays the attribution of input features (y-axis) across the 8-day input sequence (x-axis). Features within each group (Upper, Surface, Static, Convective) are sorted vertically by the absolute magnitude of their Day 5 attribution, from highest to lowest importance. The overall importance rank for each feature within its model is shown in parentheses. Red cells indicate that an increase in the feature value contributes to a higher predicted probability, while blue indicates a contribution to a lower probability. Attributions for future predictors (Days 6–8) are effectively zero due to the causal mask, with the small, non-zero IG values arising from numerical artifacts in masked-attention softmax precision and IG integration. Attributions are averaged on common cases in the test period with relatively with probability threshold exceeding 0.2 on Day 5 for DOT(GFS+Convect), DOT(GFS), DOT(Pangu-GFS) and DOT(Pangu-HRES). See supplementary material Fig. S3 for other models and days.}\label{fig: xai}
\end{figure}

Figure \ref{fig: xai} presents an example of Integrated Gradients (IG) attribution heatmaps comparing DOT(GFS+Convect) (Fig. \ref{fig: xai}a-d) and DOT(Pangu-GFS) (Fig. \ref{fig: xai}e-g) for Day 5 severe weather forecasts, with additional model and day comparisons available in the supplementary material (Fig. S3). Consistent with the transformer architecture using a causal mask, the features corresponding to the target forecast day (Day 5, the 5th column in each heatmap) generally exhibit the largest attribution magnitudes for both models. This signifies the strong reliance on the atmospheric conditions forecast for the specific target day. However, features from preceding days (Days 1-4) also show non-zero attributions, indicating that the model leverages the temporal evolution of the environment leading up to Day 5. Attributions for subsequent days (Days 6–8) are effectively zero, reflecting the causal mask that prevents the model from using future predictors. The small, non-zero IG values are due to numerical noise from masked-attention softmax precision limits, floating-point accumulation in IG integration, and the fact that inference is performed with all Days 1–8 predictors present. These artifacts are many orders of magnitude smaller than the relevant attributions and have no impact on interpretation. Lastly, static variable like the day of lead time for forecasting is ranked in the top 5 important features in both models. 

Notable differences emerge due to the distinct input feature sets, particularly the inclusion of explicit convective parameters in DOT(GFS+Convect) (panel d) which are absent in DOT(Pangu-GFS). For DOT(Pangu-GFS), which lacks convective variables, key upper-air features influencing the Day 5 prediction include humidity (q850, rank 1) and temperature (t700 rank 2, t850 rank 3), alongside surface temperature (t2m, rank 4) and shear (SHEAR850, rank 7). The signs suggest the model learns proxies for instability (e.g., higher low-level moisture/temperature, specific vertical temperature profiles) and favorable wind profiles.

In comparison, while DOT(GFS+Convect) also heavily relies on surface temperature (t2m, rank 1), it can additionally and explicitly use convective parameters like surface-based CAPE (sbcape, rank 3) which shows a strong positive attribution on Day 5. Interestingly, 2m specific humidity (q2m, rank 6) shows a counterintuitive pattern: positive attribution in the days leading up to Day 5, switching to negative attribution on Day 5 itself, potentially reflecting model behavior related to potential ongoing convection. While convective parameters like SBCAPE and precipitation are, unsurprisingly, important in GFS+Convect, 0-3km storm relative helicity (SRH03) exhibits less influence than expected based on prior studies (e.g. \cite{kerr1996storm}) linking it to supercell risk. A consistent, yet counterintuitive, pattern emerges in models for features such as q2m, z200, and u850 where the positive attribution preceding the forecast day shifts to the negative attribution on the forecast day. This attribution behavior is consistent across most of the forecast models and lead-times. For u850, stronger negative zonal winds may indicate more strongly backed flow or better moisture advection on the day of a severe weather event, but in the days prior, a more westerly zonal wind may be more conducive for severe weather during the following days. Other fields, such as q2m in the GFS model have a similar behavior, and may reflect systematic variations in moisture due to the influence of ongoing convection in the model during the forecast day.
 
Further investigation, including detailed meteorological analysis of cases exhibiting this sign switch and potentially sensitivity experiments within the models, would be necessary to fully elucidate the underlying mechanisms driving these attribution patterns.

\section{Case Studies}
Here we review several cases where the differences between the DOT(Pangu-GFS) and DOT(GFS) forecasts were most notable. To identify these events, we computed daily BSS values for each forecast where the total number of 80-km grid boxes containing a storm report exceeded 25 for the day. In doing so, we focused on the differences of the forecasts when substantial severe weather was observed within the CONUS and not on differences in false alarms between the two forecast datasets. Rather than using the severe weather report climatology as the BSS reference, the BSS for the DOT(Pangu-GFS) forecast was computed using the DOT(GFS) forecast as the reference. In this case, positive BSS values indicated that DOT(Pangu-GFS) was skillful relative to the DOT(GFS), and can be interpreted as a percent increase or decrease in the Brier Score between the two forecasts. To further distill the scores and identify events where the two hazard forecasts possessed large differences, the BSSs were computed by aggregating all 8 forecasts valid for each forecast day from each dataset.

\begin{figure}[t]
 \centering\noindent\includegraphics[width=\textwidth,angle=0]{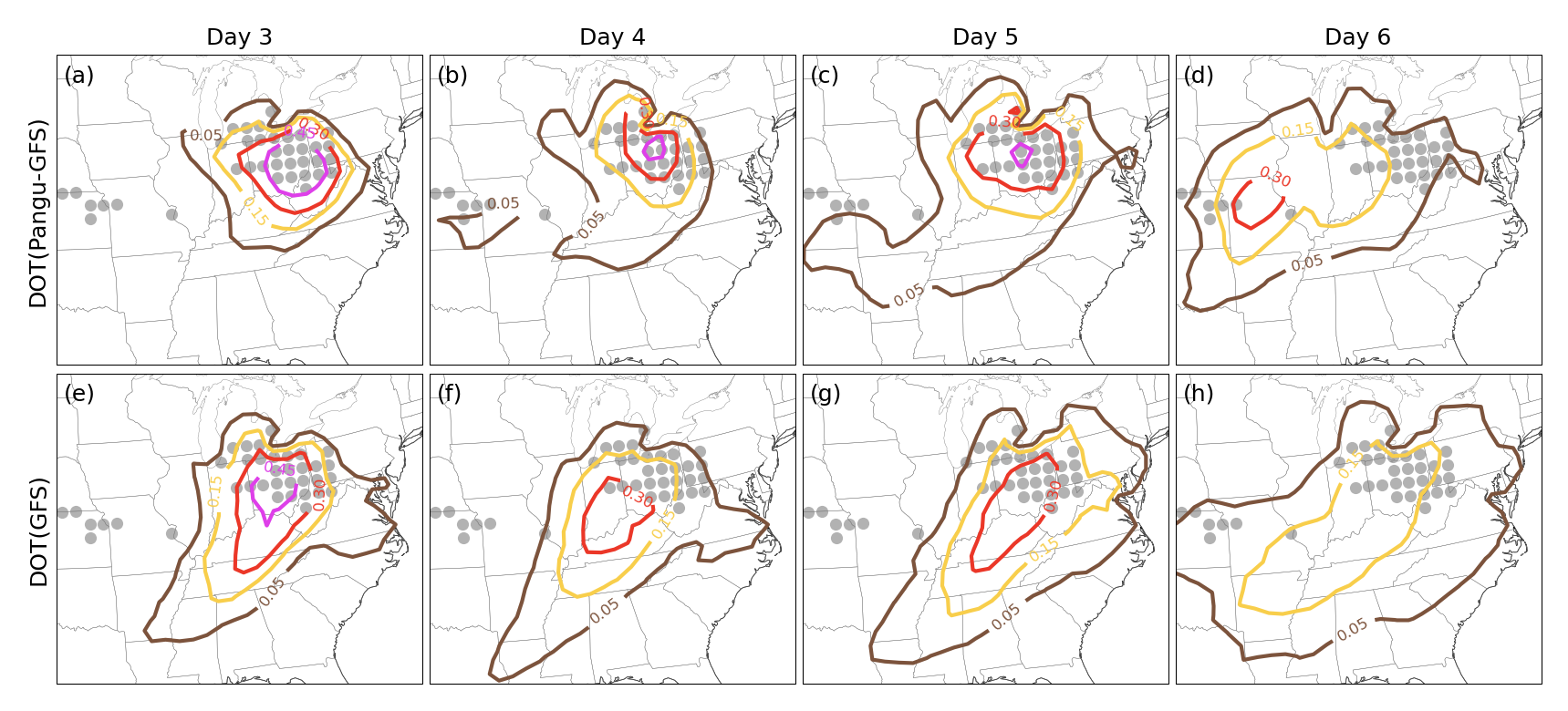}\\
 \caption{(a-d) DOT(Pangu-GFS) and (e-h) DOT(GFS) forecasts valid 12 UTC 17 April 2024 -- 12 UTC 18 April 2024 for lead-times of (a,e) 3, (b,f) 4, (c,g) 5, and (d,h) 6 days. Grid boxes containing $\geq$ 1 storm report are shown as gray circles. Forecast probabilities of 5, 15, 30, and 45\% are contoured. }
 \label{fig:case17april} 
\end{figure}

\begin{figure}[t]
 \centering\noindent\includegraphics[width=\textwidth,angle=0]{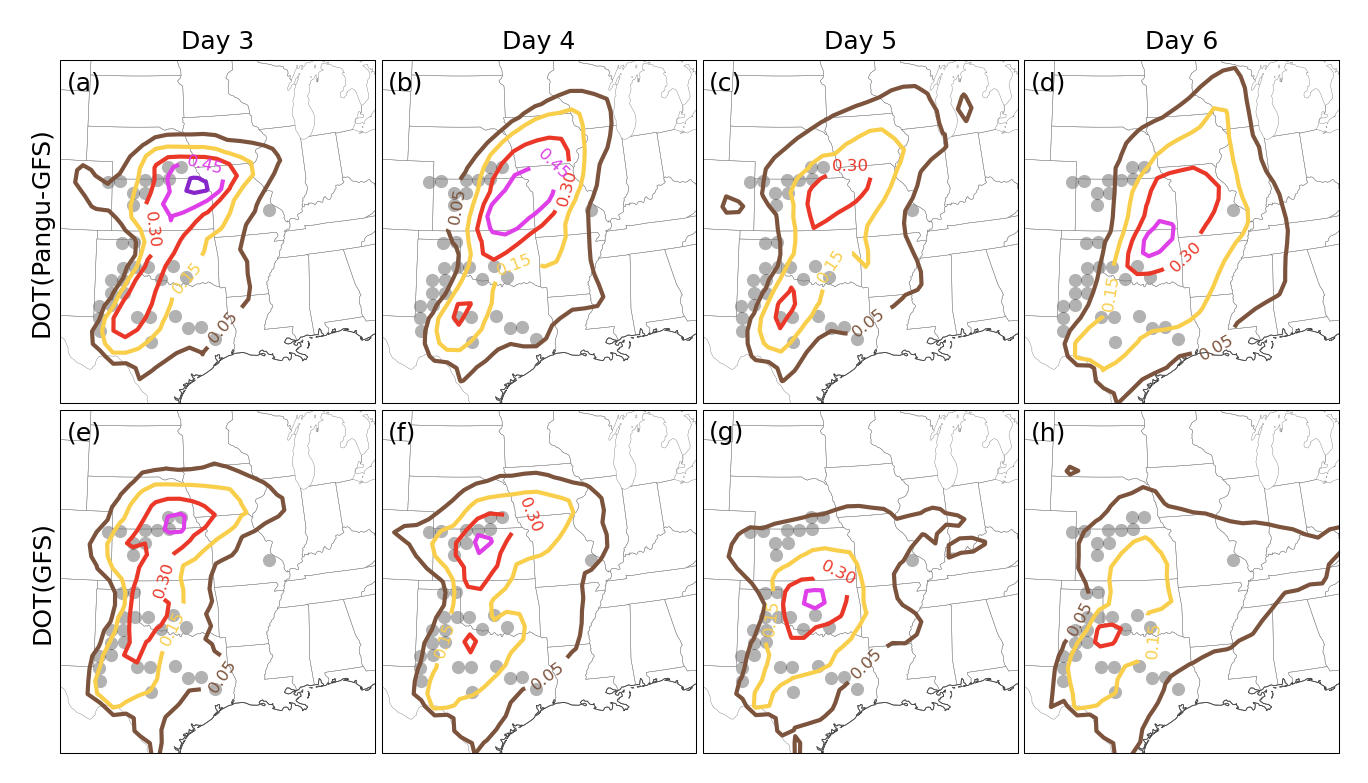}\\
 \caption{As in Fig. \ref{fig:case17april}, but for forecasts valid 12 UTC 1 May 2024 -- 12 UTC 2 May 2024. }\label{fig:case1may}
\end{figure}

The top 10 largest positive and negative BSSs are provided in Table \ref{bss-table}, along with the number of 80-km grid boxes containing $\geq$ 1 storm report for each event. We present two cases here from the Top 5 BSSs: 17 April 2024 (BSS of 0.12; Fig. \ref{fig:case17april}) and 1 May 2024 (BSS of -0.16; Fig. \ref{fig:case1may}). The 17 April 2024 event had the second highest aggregate BSS of all cases and is an example where DOT(Pangu-GFS) provided excellent medium-range guidance for a spatially compact severe weather event. The DOT(Pangu-GFS) forecasts were superior to the DOT(GFS) forecasts for the Day 4--5 lead-times (c.f., Figs. \ref{fig:case17april}b,f and Fig. \ref{fig:case17april}c,g), capturing the focused area of severe weather reports across MI, OH, and PA up to 5 days in advance and providing consistent guidance between Day 3--5. At lead-times of 4--6 days, the maximum hazard forecast probabilities within the DOT(GFS) forecast were shifted the south (Fig. \ref{fig:case17april}f--h), for the Day 4--6 forecasts, eventually focusing on the correct area by Day 3 (Fig. \ref{fig:case17april}e).

\begin{table}
\centering
\small
\begin{tabular}{ m{15em} m{15em} } 
 \hline\hline
 BSS $>$ 0 & BSS $<$ 0 \\
 \hline
28 April 2024 (0.14; 26) & 1 May 2024 (-0.16; 34) \\
16 April 2024 (0.14; 48) & 11 April 2024 (-0.11; 32) \\
17 April 2024 (0.12; 37) & 30 June 2024 (-0.10; 74) \\
18 April 2024 (0.12; 34) & 25 July 2024 (-0.05; 30) \\
16 May 2024 (0.11; 41) & 2 July 2024 (-0.05; 26) \\
14 March 2024 (0.10; 97) & 23 May 2024 (-0.04; 101) \\
26 May 2024 (0.10; 152) & 2 May 2024 (-0.04; 28) \\
10 July 2024 (0.09; 46) & 4 March 2024 (-0.04; 25) \\
30 July 2024 (0.09; 74) & 17 May 2024 (-0.04; 49) \\
27 February 2024 (0.08; 33) & 8 June 2024 (-0.04; 34) \\

\hline
\end{tabular}
\caption{Event dates with the 10 largest positive and negative BSSs for DOT(Pangu-GFS) using DOT(GFS) as the reference forecast for events where $\geq$ 25 grid boxes contained $\geq$ 1 storm report. BSSs were aggregated across all 8 forecast lead-times for each valid date. BSS values and number of grid boxes containing $\geq$ 1 storm report are provided in parentheses.}
\label{bss-table}
\end{table}

The 1 May 2024 event had the largest negative BSS, indicating the most significant difference in aggregate Brier Scores between all pairs of forecasts in favor of the DOT(GFS) forecasts. In this case, DOT(Pangu-GFS) produced large areas of false alarms over KS and MO, with forecast probabilities shifted too far to the east relative to the corridor of severe weather reports across western KS, OK, and TX (Fig. \ref{fig:case1may}a--d). This is especially notable over west TX, where the Day 4--6 DOT(Pangu-GFS) forecasts had probabilities $<$ 5\% where reports occurred (Fig. \ref{fig:case1may}b--d). In this event, severe convection was initiated by a dryline (not shown), a feature that may not have been adequately represented in the underlying Pangu-GFS forecasts compared to the GFS. By Day 3, the differences between the forecasts were less pronounced, although an eastward bias was still present in DOT(Pangu-GFS) (c.f., Fig. \ref{fig:case1may}a,e).

\section{Summary and Discussion}
This study introduces a novel approach to medium-range severe weather prediction by using Pangu-Weather, an AI NWP emulator, to generate medium-range global weather forecasts and a decoder-only transformer (DOT) architecture for post-processing the AI NWP forecasts to produce convective hazard probabilities over the CONUS. We examined the skill of a variety of hazard forecasts, including a baseline that uses the operational GFS, and three Pangu-Weather forecasts initialized with GFS, ECMWF HRES, and ERA5 initial conditions. We also directly compared the DOT against a simpler dense neural network (DNN), which was used in prior work. The transformer approach treats forecast lead times as sequential tokens, enabling the transformer to learn complex temporal relationships in the evolving atmospheric state between forecast days.

The transformer-based approach consistently outperformed traditional DNN across all forecast metrics and lead times. This performance improvement was particularly pronounced in the medium range (Days 3-6), demonstrating the value of leveraging temporal dependencies within the forecast sequence. Hazard forecasts using Pangu-Weather initialized with the GFS initial conditions (Pangu-GFS) outperformed forecasts based on the operational GFS between Days 3--7. Pangu-Weather initialized with ERA5 analyses (Pangu-ERA5) also outperformed the Pangu-GFS. This is likely due to the fact that Pangu-Weather was trained with ERA5 analyses, in addition to the more accurate ERA5 initial condition.

Further, combining multiple models into an ensemble produced the most robust forecasts, with the operational model mean consistently achieving the highest skill across all metrics and forecast days. This highlights the potential for integrating diverse model outputs to enhance operational forecast performance and mitigate individual model weaknesses. Feature attribution analysis using Integrated Gradients showed that even without explicit convective parameters, Pangu-Weather was able to leverage information from standard meteorological variables to effectively predict severe weather across multiple forecast days. However, limitation of our current attribution visualization and analysis is the temporal aggregation across 24-hour periods, which may obscure important subdaily variations in feature importance. Future investigations should consider implementing subdaily attribution analysis, similar to the tree interpreter approach of \cite{mazurek2025can}, to better understand the temporal evolution of feature contributions and validate the meteorological reasonableness of observed attribution patterns.

Case studies revealed both strengths and limitations of AI-based forecasts. DOT(Pangu-GFS) excelled at predicting spatially compact severe weather events several days in advance, it occasionally produced eastward-biased forecasts and struggled with certain meteorological phenomena like drylines. Additionally, reliability analysis showed that individual models tend to slightly over-forecast at higher probability thresholds, though this tendency diminishes at longer lead times. While the AI models demonstrated promising skill during the unusually active 2024 season, a longer verification period spanning multiple years is necessary to fully assess the generalizability of these results and quantify skill variations across different climatological regimes.

Future work should explore incorporating a wider variety of AI models, developing methods to derive convective parameters directly from AI model outputs, and further investigating the meteorological factors or non meteorological factors like elevation variability \cite{hua2019empirical} and vegetation \cite{kellner2014land} driving model performance differences. Understanding how errors and biases in the underlying AI models propagate through post-processing is key to improving these systems. Nevertheless, the framework presented here offers a promising pathway for operational implementation of AI-enhanced medium-range severe weather prediction, potentially improving both forecast accuracy and lead time for high-impact events.

\section*{Acknowledgements}
Support for this work was provided by the NSF NCAR Short-term Explicit Prediction program, the NSF NCAR Computational and Information Sciences Laboratory visitor fund, and NSF PREEVENTS award number 1854966. ZH and AAF acknowledge support from the National Science Foundation under Grant No. 2209699.

\section*{Data availability statement}
NWS Storm Reports that were used for model training were retrieved from the NOAA SPC storm report archive located at https://www.spc.noaa.gov/wcm/\#data. The Pangu-Weather model is available at https://github.com/198808xc/Pangu-Weather. Initial conditions, including ERA5, HRES, and GFS analyses were retrieved from the NCAR Research Data Archive located at https://rda.ucar.edu. The code implementation and model checkpoints will be available on the Zenodo repository upon acceptance of this work.

\bibliographystyle{ametsocV6.bst}  
\bibliography{references}  

\begin{thebibliography}{42}
\providecommand{\natexlab}[1]{#1}
\providecommand{\url}[1]{\texttt{#1}}
\renewcommand{\UrlFont}{\rmfamily}
\providecommand{\urlprefix}{URL }
\expandafter\ifx\csname urlstyle\endcsname\relax
  \providecommand{\doi}[1]{https://doi.org/\discretionary{}{}{}#1}\else
  \providecommand{\doi}{https://doi.org/\discretionary{}{}{}\begingroup \urlstyle{rm}\Url}\fi
\providecommand{\eprint}[2][]{\url{#2}}

\bibitem[{Bi et~al.(2023)Bi, Xie, Zhang, Chen, Gu,, and Tian}]{bi2023accurate}
Bi, K., L.~Xie, H.~Zhang, X.~Chen, X.~Gu, and Q.~Tian, 2023: {Accurate} {Medium}-{Range} {Global} {Weather} {Forecasting} with 3{D} {Neural} {Networks}. \textit{Nature}, \textbf{619~(7970)}, 533--538.

\bibitem[{Bouall{\`e}gue et~al.(2024)Bouall{\`e}gue, Weyn, Clare, Dramsch, Dueben,, and Chantry}]{bouallegue2024improving}
Bouall{\`e}gue, Z.~B., J.~A. Weyn, M.~C. Clare, J.~Dramsch, P.~Dueben, and M.~Chantry, 2024: Improving medium-range ensemble weather forecasts with hierarchical ensemble transformers. \textit{Artificial Intelligence for the Earth Systems}, \textbf{3~(1)}, e230\,027.

\bibitem[{Bremnes et~al.(2024)Bremnes, Nipen,, and Seierstad}]{npg-31-247-2024}
Bremnes, J.~B., T.~N. Nipen, and I.~A. Seierstad, 2024: Evaluation of forecasts by a global data-driven weather model with and without probabilistic post-processing at norwegian stations. \textit{Nonlinear Processes in Geophysics}, \textbf{31~(2)}, 247--257, \doi{10.5194/npg-31-247-2024}, \urlprefix\url{https://npg.copernicus.org/articles/31/247/2024/}.

\bibitem[{Bülte et~al.(2025)Bülte, Horat, Quinting,, and Lerch}]{buelte24}
Bülte, C., N.~Horat, J.~Quinting, and S.~Lerch, 2025: Uncertainty quantification for data-driven weather models. \textit{Artificial Intelligence for the Earth Systems}, \doi{10.1175/AIES-D-24-0049.1}, \urlprefix\url{https://journals.ametsoc.org/view/journals/aies/aop/AIES-D-24-0049.1/AIES-D-24-0049.1.xml}.

\bibitem[{DeMaria et~al.(2024)DeMaria, Franklin, Chirokova, Radford, DeMaria, Musgrave,, and Ebert-Uphoff}]{demaria2024evaluation}
DeMaria, M., J.~L. Franklin, G.~Chirokova, J.~Radford, R.~DeMaria, K.~D. Musgrave, and I.~Ebert-Uphoff, 2024: Evaluation of tropical cyclone track and intensity forecasts from artificial intelligence weather prediction (aiwp) models. \textit{arXiv preprint arXiv:2409.06735}.

\bibitem[{Dowell et~al.(2022)}]{Dowell2022}
Dowell, D.~C., and Coauthors, 2022: The high-resolution rapid refresh (hrrr): An hourly updating convection-allowing forecast model. part i: Motivation and system description. \textit{Wea. and Forecasting.}, \textbf{37~(8)}, 1371–1395, \doi{10.1175/WAF-D-21-0151.1}.

\bibitem[{Gagne et~al.(2019)Gagne, Haupt, Nychka,, and Thompson}]{Gagne2019-it}
Gagne, D.~J., II, S.~E. Haupt, D.~W. Nychka, and G.~Thompson, 2019: Interpretable deep learning for spatial analysis of severe hailstorms. \textit{Mon. Weather Rev.}, \textbf{147~(8)}, 2827--2845, \doi{10.1175/MWR-D-18-0316.1}, \urlprefix\url{https://journals.ametsoc.org/view/journals/mwre/147/8/mwr-d-18-0316.1.xml}.

\bibitem[{Hersbach et~al.(2020)}]{hersbach2020era5}
Hersbach, H., and Coauthors, 2020: {The} {ERA5} {Global} {Reanalysis}. \textit{Quarterly Journal of the Royal Meteorological Society}, \textbf{146~(730)}, 1999--2049.

\bibitem[{Hill et~al.(2020)Hill, Herman,, and Schumacher}]{hill2020}
Hill, A.~J., G.~R. Herman, and R.~S. Schumacher, 2020: Forecasting severe weather with random forests. \textit{Monthly Weather Review}, \textbf{148~(5)}, 2135--2161.

\bibitem[{Hill et~al.(2023)Hill, Schumacher,, and Jirak}]{hill2023}
Hill, A.~J., R.~S. Schumacher, and I.~L. Jirak, 2023: A new paradigm for medium-range severe weather forecasts: Probabilistic random forest–based predictions. \textit{Wea. Forecasting}, \textbf{38~(2)}, 251--272.

\bibitem[{Hua and Anderson-Frey(2023)Hua, and Anderson-Frey}]{hua2023tornadic}
Hua, Z., and A.~Anderson-Frey, 2023: {How} {Are} {Tornadic} {Supercell} {Soundings} {Significantly} {Different} {From} {Nearby} {Baseline} {Environments}? \textit{Geophysical Research Letters}, \textbf{50~(8)}, e2022GL102\,580.

\bibitem[{Hua and Anderson-Frey(2022)Hua, and Anderson-Frey}]{hua2022self}
Hua, Z., and A.~K. Anderson-Frey, 2022: {Self}-{Organizing} {Maps} for the {Classification} of {Spatial} and {Temporal} {Variability} of {Tornado}-{Favorable} {Parameters}. \textit{Monthly Weather Review}, \textbf{150~(2)}, 393--407.

\bibitem[{Hua and Chavas(2019)Hua, and Chavas}]{hua2019empirical}
Hua, Z., and D.~R. Chavas, 2019: The empirical dependence of tornadogenesis on elevation roughness: Historical record analysis using bayes’s law in arkansas. \textit{Journal of Applied Meteorology and Climatology}, \textbf{58~(2)}, 401--411.

\bibitem[{Hua et~al.(2025)Hua, Hakim,, and Anderson-Frey}]{Hua2025performance}
Hua, Z., G.~Hakim, and A.~Anderson-Frey, 2025: Performance of the pangu-weather deep learning model in forecasting tornadic environments. \textit{Geophysical Research Letters}, \textbf{52~(7)}, e2024GL109\,611.

\bibitem[{Kellner and Niyogi(2014)Kellner, and Niyogi}]{kellner2014land}
Kellner, O., and D.~Niyogi, 2014: Land surface heterogeneity signature in tornado climatology? an illustrative analysis over indiana, 1950--2012. \textit{Earth Interactions}, \textbf{18~(10)}, 1--32.

\bibitem[{Kerr and Darkow(1996)Kerr, and Darkow}]{kerr1996storm}
Kerr, B.~W., and G.~L. Darkow, 1996: Storm-relative winds and helicity in the tornadic thunderstorm environment. \textit{Weather and forecasting}, \textbf{11~(4)}, 489--505.

\bibitem[{Li et~al.(2024)Li, Geiss, Feng, Leung, Qian,, and Cui}]{li2024derecho}
Li, J., A.~Geiss, Z.~Feng, L.~R. Leung, Y.~Qian, and W.~Cui, 2024: A derecho climatology (2004--2021) in the united states based on machine learning identification of bow echoes. \textit{Earth System Science Data Discussions}, \textbf{2024}, 1--41.

\bibitem[{Loken et~al.(2020)Loken, Clark,, and Karstens}]{loken2020}
Loken, E.~D., A.~J. Clark, and C.~D. Karstens, 2020: Generating probabilistic next-day severe weather forecasts from convection-allowing ensembles using random forests. \textit{Weather and Forecasting}, \textbf{35~(4)}, 1605--1631.

\bibitem[{Loken et~al.(2022)Loken, Clark,, and McGovern}]{loken2022}
Loken, E.~D., A.~J. Clark, and A.~McGovern, 2022: Comparing and interpreting differently designed random forests for next-day severe weather hazard prediction. \textit{Weather and Forecasting}, \textbf{37~(6)}, 871--899.

\bibitem[{Mazurek et~al.(2025)Mazurek, Hill, Schumacher,, and McDaniel}]{mazurek2025can}
Mazurek, A.~C., A.~J. Hill, R.~S. Schumacher, and H.~J. McDaniel, 2025: Can ingredients-based forecasting be learned? disentangling a random forest’s severe weather predictions. \textit{Weather and Forecasting}, \textbf{40~(2)}, 237--258.

\bibitem[{McGovern et~al.(2023)McGovern, Chase, Flora, Gagne~II, Lagerquist, Potvin, Snook,, and Loken}]{mcgovern2023}
McGovern, A., R.~J. Chase, M.~Flora, D.~J. Gagne~II, R.~Lagerquist, C.~K. Potvin, N.~Snook, and E.~Loken, 2023: A review of machine learning for convective weather. \textit{Artificial Intelligence for the Earth Systems}, \textbf{2~(3)}, \doi{10.1175/AIES-D-22-0077.1}.

\bibitem[{Meng et~al.(2025)Meng, Hakim, Yang,, and Vecchi}]{meng2025deep}
Meng, Z., G.~J. Hakim, W.~Yang, and G.~A. Vecchi, 2025: Deep learning atmospheric models reliably simulate out-of-sample land heat and cold wave frequencies. \textit{arXiv preprint arXiv:2507.03176}.

\bibitem[{Paszke(2019)}]{paszke2019pytorch}
Paszke, A., 2019: Pytorch: An imperative style, high-performance deep learning library. \textit{arXiv preprint arXiv:1912.01703}.

\bibitem[{Radford et~al.(2019)Radford, Wu, Child, Luan, Amodei, Sutskever et~al.}]{radford2019language}
Radford, A., J.~Wu, R.~Child, D.~Luan, D.~Amodei, I.~Sutskever, and Coauthors, 2019: Language models are unsupervised multitask learners. \textit{OpenAI blog}, \textbf{1~(8)}, 9.

\bibitem[{Rasp et~al.(2023)}]{rasp2023weatherbench}
Rasp, S., and Coauthors, 2023: {Weatherbench} 2: {A} {Benchmark} for the {Next} {Generation} of {Data}-{Driven} {Global} {Weather} {Models}. \textit{arXiv preprint arXiv:2308.15560}.

\bibitem[{Schulz and Lerch(2022)Schulz, and Lerch}]{schulz2022machine}
Schulz, B., and S.~Lerch, 2022: Machine learning methods for postprocessing ensemble forecasts of wind gusts: A systematic comparison. \textit{Monthly Weather Review}, \textbf{150~(1)}, 235--257.

\bibitem[{Sha et~al.(2024)Sha, Sobash,, and Gagne}]{Shaetal2024}
Sha, Y., R.~A. Sobash, and D.~J. Gagne, 2024: Generative ensemble deep learning severe weather prediction from a deterministic convection-allowing model. \textit{Artificial Intelligence for the Earth Systems}, \textbf{3~(2)}, e230\,094, \doi{10.1175/AIES-D-23-0094.1}, \urlprefix\url{https://journals.ametsoc.org/view/journals/aies/3/2/AIES-D-23-0094.1.xml}.

\bibitem[{Shield and Houston(2022)Shield, and Houston}]{shield2022diagnosing}
Shield, S.~A., and A.~L. Houston, 2022: Diagnosing supercell environments: A machine learning approach. \textit{Weather and Forecasting}, \textbf{37~(5)}, 771--785.

\bibitem[{Sobash and Ahijevych(2024)Sobash, and Ahijevych}]{SobashAhijevych2024}
Sobash, R.~A., and D.~Ahijevych, 2024: Evaluating machine learning-based probabilistic convective hazard forecasts using the {HRRR}: Quantifying hazard predictability and sensitivity to training choices. \textit{Wea. Forecasting}, \doi{10.1175/WAF-D-23-0221.1}, \urlprefix\url{https://journals.ametsoc.org/view/journals/wefo/aop/WAF-D-23-0221.1/WAF-D-23-0221.1.xml}.

\bibitem[{Sobash and Ahijevych(2025)Sobash, and Ahijevych}]{SobashAhijevych2025}
Sobash, R.~A., and D.~Ahijevych, 2025: Evaluating machine learning-based probabilistic lightning forecasts using the hrrr: A comparison to three forecast baselines. \textit{Wea. Forecasting}, \urlprefix\url{in press}.

\bibitem[{Sobash et~al.(2020)Sobash, Romine,, and Schwartz}]{Sobash2020-zs}
Sobash, R.~A., G.~S. Romine, and C.~S. Schwartz, 2020: A comparison of {Neural-Network} and {Surrogate-Severe} probabilistic convective hazard guidance derived from a {Convection-Allowing} model. \textit{Wea. Forecasting}, \textbf{35~(5)}, 1981--2000, \doi{10.1175/WAF-D-20-0036.1}, \urlprefix\url{https://journals.ametsoc.org/view/journals/wefo/35/5/wafD200036.xml}.

\bibitem[{SPC(2025)}]{SPCwcm}
SPC, 2025: {SPC} severe weather climatology. Storm Prediction Center, accessed: 2025-02-26, \url{https://www.spc.noaa.gov/wcm/}.

\bibitem[{Sundararajan et~al.(2017)Sundararajan, Taly,, and Yan}]{sundararajan2017axiomatic}
Sundararajan, M., A.~Taly, and Q.~Yan, 2017: Axiomatic attribution for deep networks. \textit{International conference on machine learning}, PMLR, 3319--3328.

\bibitem[{Thompson et~al.(2003)Thompson, Edwards, Hart, Elmore,, and Markowski}]{Thompson2003-co}
Thompson, R.~L., R.~Edwards, J.~A. Hart, K.~L. Elmore, and P.~Markowski, 2003: Close proximity soundings within supercell environments obtained from the rapid update cycle. \textit{Wea. Forecasting}, \textbf{18~(6)}, 1243--1261, \doi{10.1175/1520-0434(2003)018<1243:CPSWSE>2.0.CO;2}, \urlprefix\url{https://journals.ametsoc.org/view/journals/wefo/18/6/1520-0434_2003_018_1243_cpswse_2_0_co_2.xml}.

\bibitem[{Thompson et~al.(2013)Thompson, Smith, Dean,, and Marsh}]{Thompson2013-wz}
Thompson, R.~L., B.~T. Smith, A.~R. Dean, and P.~T. Marsh, 2013: Spatial distributions of tornadic near-storm environments by convective mode. \textit{E-Journal of Severe Storms Meteorology}, \textbf{8~(5)}, \urlprefix\url{https://www.spc.noaa.gov/publications/thompson/ej-env.pdf}.

\bibitem[{Van~Poecke et~al.(2025)}]{van2025self}
Van~Poecke, A., and Coauthors, 2025: Self-attentive transformer for fast and accurate postprocessing of temperature and wind speed forecasts. \textit{Artificial Intelligence for the Earth Systems}, \textbf{1~(aop)}.

\bibitem[{Vannitsem et~al.(2021)}]{vannitsem2021statistical}
Vannitsem, S., and Coauthors, 2021: Statistical postprocessing for weather forecasts: Review, challenges, and avenues in a big data world. \textit{Bulletin of the American Meteorological Society}, \textbf{102~(3)}, E681--E699.

\bibitem[{Vaswani(2017)}]{vaswani2017attention}
Vaswani, A., 2017: Attention is all you need. \textit{Advances in Neural Information Processing Systems}.

\bibitem[{Veldkamp et~al.(2021)Veldkamp, Whan, Dirksen,, and Schmeits}]{veldkamp2021statistical}
Veldkamp, S., K.~Whan, S.~Dirksen, and M.~Schmeits, 2021: Statistical postprocessing of wind speed forecasts using convolutional neural networks. \textit{Monthly Weather Review}, \textbf{149~(4)}, 1141--1152.

\bibitem[{Wilks(2011)}]{wilks2011statistical}
Wilks, D.~S., 2011: \textit{Statistical methods in the atmospheric sciences}. Academic press.

\bibitem[{Yang et~al.(2006)Yang, Pan, Krueger, Moorthi,, and Lord}]{yang2006evaluation}
Yang, F., H.-L. Pan, S.~K. Krueger, S.~Moorthi, and S.~J. Lord, 2006: {Evaluation} of the {NCEP} {Global} {Forecast} {System} at the {ARM} {SGP} {Site}. \textit{Monthly Weather Review}, \textbf{134~(12)}, 3668--3690.

\bibitem[{Zhuang(2019)}]{zhuang2019xesmf}
Zhuang, J., 2019: x{ESMF} {Documentation}.

\end{thebibliography}






\newpage
\appendix

\section*{Supporting Information} 
\renewcommand{\thesection}{S\arabic{section}} 
\setcounter{section}{0} 

\setcounter{figure}{0}
\renewcommand{\thefigure}{S\arabic{figure}}
\setcounter{table}{0}
\renewcommand{\thetable}{S\arabic{table}}


\begin{figure}[h]
 \includegraphics[width=37pc,angle=0]{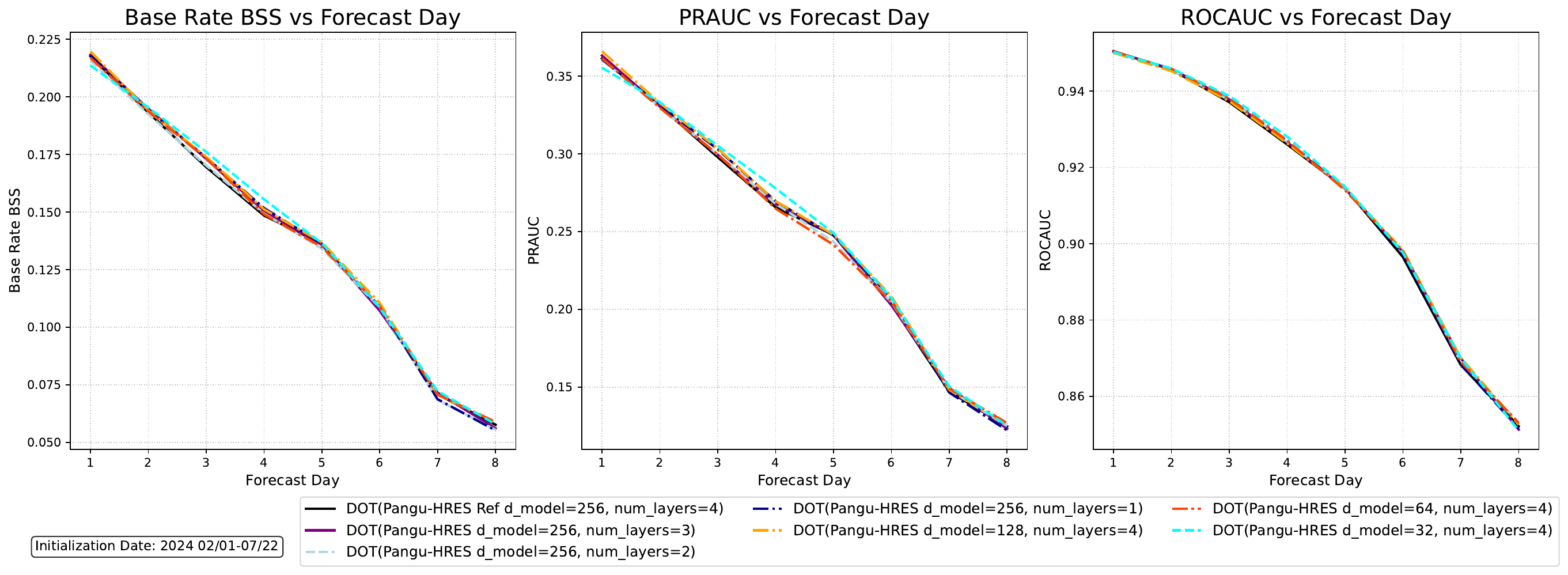}\\
 \caption{Similar to Fig. 3 but for experiment using DOT(Pangu-HRES) by either reducing ``d\_model'' from the 256 in the study down to 128, 64 and 32 or reducing ``num\_layers'' from 4 down to 3,2 and 1. No significance difference in performance with respect to the reference configuration in the formal analysis.}\label{fig: xai}
\end{figure}

\begin{figure}[h]
 \centering\noindent\includegraphics[width=37pc,angle=0]{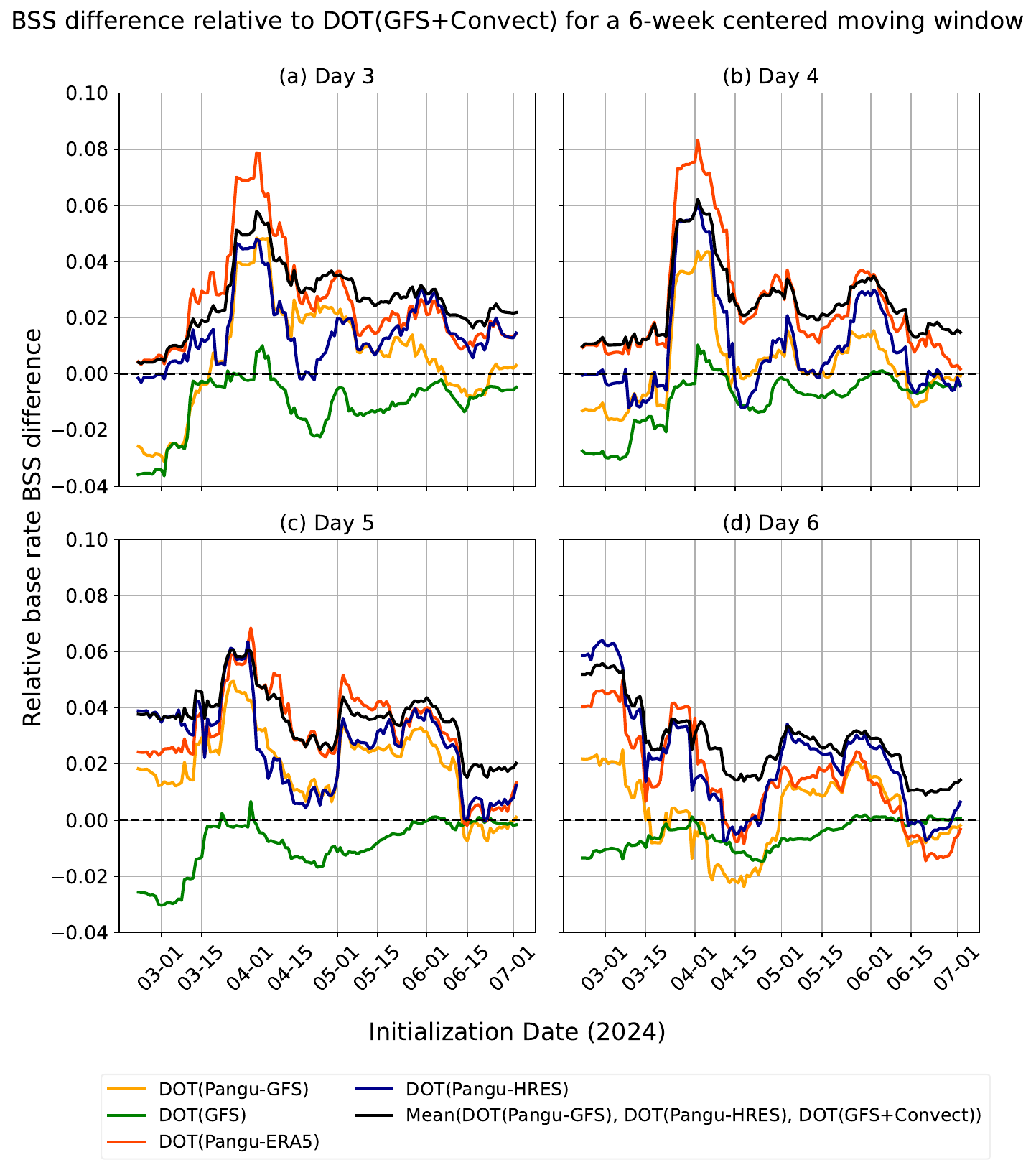}\\
 \caption{Similar to Fig. 7 but use DOT(GFS+Convect) as comparison baseline}\label{fig: xai}
\end{figure}

\begin{figure}[h]
 \centering\noindent\includegraphics[width=37pc,angle=0]{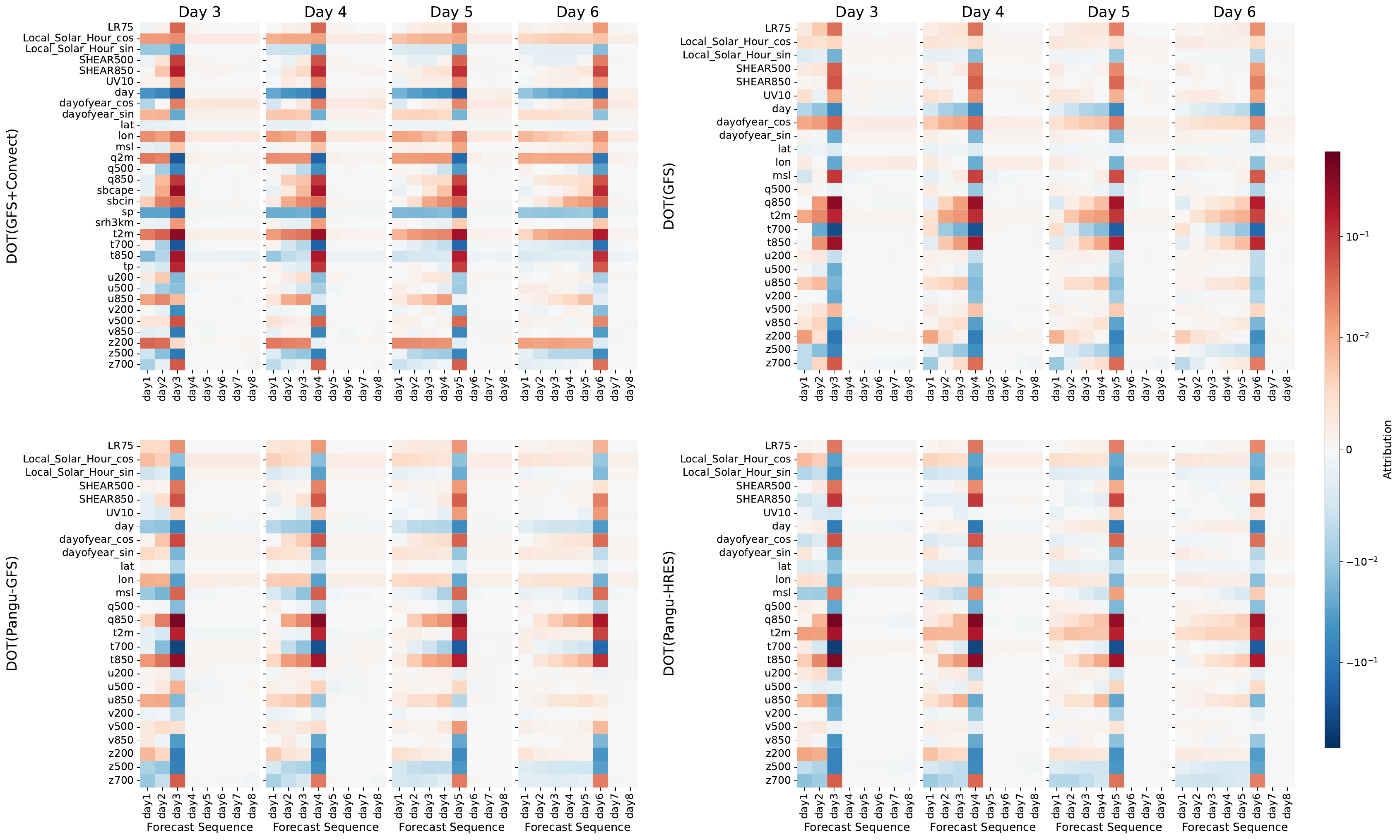}\\
 \caption{Integrated Gradients (IG) attribution heatmaps for four weather models (DOT(GFS+Convect), DOT(GFS), DOT(Pangu-GFS), DOT(Pangu-HRES)) predicting severe weather probability for forecast Days 3-6. Heatmaps show the relative importance of each input feature (y-axis) at each day in the 8-day input sequence (x-axis). Red indicates that an increase in the feature value (relative to a zero baseline) increases the predicted probability, while blue indicates a decrease. Analysis is focused on common cases in the test period with relatively large predicted probabilities exceeding thresholds describe in Section 2.c.3}\label{fig: xai}
\end{figure}

\end{document}